\documentclass[a4paper,10pt,twoside]{cpc-hepnp}

\usepackage{multicol}
\usepackage{graphicx}
\usepackage{booktabs}
\usepackage{amssymb,bm,mathrsfs,bbm,amscd}
\usepackage[tbtags]{amsmath}
\usepackage{lastpage}
\usepackage{array,multirow,graphicx}

\begin{document}

\fancyhead[c]{\small Chinese Physics C~~~Vol. xx, No. x (201x) xxxxxx}
\fancyfoot[C]{\small 010201-\thepage}

\footnotetext[0]{}

\title{Excited state mass spectra, Decay properties and Regge trajectories of charm and charm-strange mesons}

\author{%
    Virendrasinh Kher $^{1}$,$^{2}$\email{vhkher@gmail.com}%
\quad Nayneshkumar Devlani $^{1}$\email{nayneshdev@gmail.com}%
\quad Ajay Kumar Rai$^{2}$*\email{raiajayk@gmail.com}%
}
\maketitle

\address{%
$^1$ Applied Physics Department, Polytechnic, The M.S. University of Baroda, Vadodara 390002, Gujarat, India \\
$^2$ Department of Applied Physics, Sardar Vallabhbhai National Institute of Technology, Surat 395007, Gujarat, India \\
}

\begin{abstract}
The framework of a phenomenological quark-antiquark potential (Coulomb plus linear confinement) model with a Gaussian wave function is used for detailed study of masses of the ground, orbitally and radially excited states of heavy-light $Q\overline{q}$, $\left(Q=c,q=u/d,s\right)$ mesons. We incorporate a $\mathcal{O}(1/m)$ correction to the potential energy term and relativistic corrections to the kinetic energy term of the Hamiltonian. The spin-hyperfine, spin-orbit and tensor interactions incorporating the effect of mixing are employed to obtain the pseudoscalar, vector, radially and orbitally excited state meson masses. The Regge trajectories in the $(J,M^{2})$ and $(n_r,M^{2})$ planes for heavy-light mesons are investigated with their corresponding parameters.  Leptonic and radiative leptonic decay widths and corresponding branching ratios are computed. The mixing parameters are also estimated. Our predictions are in good agreement with experimental results as well as lattice and other theoretical models.
\end{abstract}



\footnotetext[0]{\hspace*{-3mm}\raisebox{0.3ex}{$\scriptstyle\copyright$}2013
Chinese Physical Society and the Institute of High Energy Physics
of the Chinese Academy of Sciences and the Institute
of Modern Physics of the Chinese Academy of Sciences and IOP Publishing Ltd}%

\begin{multicols}{2}

\section{Introduction}
\label{intro}

The discovery of many new excited states of $D$ and $D_S$ mesons through experiments by BABAR, BELLE and CLEO have ignited interest in the properties of these mesons \cite{PDGlatest}. Many theoretical models have made spectroscopic assignments to these states. Quite recently, LHCb has been able to discover and measure the properties of these excited states \cite{Aaij2015,Aaij2015a,Aaij2013,Aaij2012a,Aaij2014}. This has re-ignited interest in the spectroscopy of these heavy-light mesons, as one expects that with the availability of higher energy and luminosity beams at the LHC and higher luminosity at the SuperKEKB $e^{+}e^{-}$ − collider, more new states will be observed \cite{Godfrey2015}. There are certainly open questions regarding the nature of several orbitally excited states. For example, the $D_{S0}^{\star}(2317)$ and $D_{S1}(2460)$  are predicted to be heavier than experimentally measured \cite{PDGlatest,Chen:2016}. This indicates that these may not fit into conventional quark anti-quark bound states but could instead be candidates for exotic states.

The interpretation of the newly observed excited states, whether as conventional $Q\bar{q}$ heavy-light quark bound states or more exotic structures, is of great significance. Their weak and electromagnetic decays often provide help for identification and assignment of the $J^{PC}$ values. Furthermore, steady progress is being done in lattice calculations, and masses for various low-lying states have already been calculated \cite{Cichy2016}.

Many theoretical schemes have been employed to understand the spectroscopic properties of the $D$ and $D_S$ mesons. The ground state $1^{3}S_{1}$ and $1^{1}S_{0}$ have been estimated quite accurately. However, for $L=1$, state disparities exist between theoretical estimates and experimental measurements. In this article, we employ a potential model based approach to understand the open charm heavy-light mesons. The validity of a potential model in the light quark sector is certainly questionable and in the case of heavy-light mesons, the presence of a light quark warrants  a relativistic treatment. 

In this paper, with a view to improving the methodology of the potential model, we incorporate corrections to the potential energy part of the Hamiltonian besides the kinetic energy part, and use a unified scheme for both mesons. We investigate the Regge trajectories in the $(J,M^{2})$ and $(n_r,M^{2})$ planes ($M$ is the mass, $J$ is the total angular momentum and $n_r$ is the radial quantum number of the meson state), which is important for illustrating the nature of current and future experimentally observed heavy-light mesons \cite{Devlani2013, Devlani2011,Ebert2010}. 

 Leptonic and radiative leptonic decay widths are calculated for $D$ and $D_s$ mesons, as well as mixing parameters for the $D$ meson, are estimated in the present scheme. The radiative leptonic decays of heavy-light mesons are important, as the strong interaction is involved only in one hadronic external state, which only occurs within the initial particle. The helicity is suppressed because this decay rate is proportional to the square of the lepton mass ($m_{l}^{2}$).  The helicity suppression can be compensated by the presence of one photon in the final state. It thus opens a window for studying the effect of strong interactions in the decay \cite{Yang:2012}.

 The article is organized as follows. We present the details of the theoretical framework for the calculation of mass spectra in Section {\ref{sec:mass}} and {\ref{sec:mass2}}, leptonic and radiative leptonic decays in Section {\ref{sec:lepto}}, and mixing parameters for the $D$ meson in Section {\ref{sec:mixing}}. Results for the  mass spectra, leptonic, and radiative leptonic decays for both the $D$ and $D_s$ meson, as well as mixing parameters for the $D$ meson, are discussed in Section {\ref{sec:Resu}}. The Regge trajectories in the $(J,M^{2})$ and $(n_r,M^{2})$ planes are in Section {\ref{sec:reg}}. Finally, we draw our conclusion in  Section-{\ref{sec:conclusion}}.

\section{Methodology}

\subsection{Cornell potential with ${\cal{O}}\left(\frac{1}{m}\right)$ corrections \label{sec:mass}}

For the study of the heavy-light mesons, we employ the following Hamiltonian \cite{Devlani2011,Devlani2012,Devlani2013}. 

\begin{equation}
H=\sqrt{\mathbf{p}^{2}+m_{Q}^{2}}+\sqrt{\mathbf{p}^{2}+m_{\bar{q}}^{2}}+V(\mathbf{r});\label{Eq:hamiltonian}
\end{equation} 
where $\mathbf{p}$ is the relative momentum of the quark-antiquark, $m_{Q}$ is the mass of the heavy quark, $m_{\bar{q}}$ is the mass of the light anti-quark, and \ensuremath{V(\mathbf{r})} is the quark-antiquark potential, which can be written as \cite{Koma2006}, 
\begin{equation}
V\left(r\right)=V^{\left(0\right)}\left(r\right)+\left(\frac{1}{m_{Q}}+\frac{1}{m_{\bar{q}}}\right)V^{\left(1\right)}\left(r\right)+{\cal O}\left(\frac{1}{m^{2}}\right)
\end{equation}.

 $V^{\left(0\right)}$ is the Cornell like potential~\cite{Eichten1978}, 
 
\begin{equation}
V^{\left(0\right)}(r)=-\frac{\alpha_{c}}{r}+Ar+V_{0}
\end{equation}
where $A$ is a potential parameter and $V_{0}$ is a constant. $\alpha_{c}=(4/3)\alpha_{S}\left({M^{2}}\right)$;
$\alpha_{S}\left({M^{2}}\right)$ is the strong running coupling constant. The non-perturbative form of $V^{\left(1\right)}\left(r\right)$ is not yet known, but leading order perturbation theory yields \begin{equation}
V^{\left(1\right)}\left(r\right)=-C_{F}C_{A}\alpha_{s}^{2}/4r^{2}
\end{equation}
where $C_{F}=4/3$ and $C_{A}=3$ are the Casimir charges of the fundamental and adjoint representation respectively \cite{Koma2006}.

For the present study, we use the Ritz variational scheme.  The confining interaction plays an important role in the heavy-light mesons. Based on the results outlined in Refs.~\cite{Devlani2013} and \cite{Devlani2011}, a Gaussian wave function is most suitable for the present study of heavy-light mesons. Thus to obtain the expectation values of the Hamiltonian we employ a Gaussian wave function. These wave functions are taken from the solution of the relativistic Hamiltonian, as described in Ref.~\cite{Danilkin2010} (and references therein). The authors of that article have used these wave functions for the study of light, heavy-light and heavy-heavy flavoured  mesons.  The Gaussian wave function in position space has the form 
\begin{eqnarray}
R_{nl}(\mu,r) & = & \mu^{3/2}\left(\frac{2\left(n-1\right)!}{\Gamma\left(n+l+1/2\right)}\right)^{1/2}\left(\mu r\right)^{l}\times\nonumber \\
 &  & e^{-\mu^{2}r^{2}/2}L_{n-1}^{l+1/2}(\mu^{2}r^{2})
\end{eqnarray} and in momentum space has the form  \begin{eqnarray}
R_{nl}(\mu,p) & = & \frac{\left(-1\right)^{n}}{\mu^{3/2}}\left(\frac{2\left(n-1\right)!}{\Gamma\left(n+l+1/2\right)}\right)^{1/2}\left(\frac{p}{\mu}\right)^{l}\times\nonumber \\
 &  & e^{-{p}^{2}/2\mu^{2}}L_{n-1}^{l+1/2}\left(\frac{p^{2}}{\mu^{2}}\right).
\end{eqnarray}
Here, $\mu$ is the variational parameter and $L$ is a Laguerre polynomial.

For a chosen value of $A$, the variational parameter $\mu$ is determined for each state using the Virial theorem~\cite{Hwang1997a},

\begin{equation}
 \left\langle{K.E.}\right\rangle =\frac{1}{2} \left\langle{\frac{rdV}{dr}}\right\rangle.
\end{equation} 

Since the quarks within the heavy-light mesons are relativistic, a non-relativistic approach is not justified. Therefore, for the K.E., we expand the kinetic energy of the quarks from the Hamiltonian equation~(1), retaining powers up to ${{\cal{O}}\left({\bf p}^{10}\right)}$, to incorporate the relativistic correction. This allows one to extend the potential model approach to heavy-light systems. 
In the series expansion of kinetic energy, for $v<c$, the effect of the higher order terms of the momentum ${\bf p}^{2n}(n > 2)$ is very small compared to ${\bf p}^2$. Even more, the higher order terms have poor convergence. The expansion term up to ${\bf p}^4$ does not have a lower bound and the powers of $\bf p$ have canceling contributions, as terms with ${\bf p}^4$ and ${\bf p}^8$ are negative whereas  ${\bf p}^6$ and ${\bf p}^{10}$ terms are positive. So, the usable expansion to incorporate the relativistic effect is up to ${{\cal{O}}\left({\bf P}^{10}\right)}$\cite{Rai2015epj}. We employ a position space Gaussian wave function to obtain the expectation value of the potential energy part in the Virial theorem, while the momentum space wave function has been used to obtain the kinetic energy part. 

Since the interaction potential does not contain spin-dependent terms, the expectation value of the Hamiltonian yields spin-averaged mass. The ground state spin-averaged mass is matched with the PDG value by fixing the potential constant $V_0$. The spin-averaged mass for the ground state is computed using the equation \cite{Rai2008},

\begin{equation}
M_{SA}=M_{P}+\frac{3}{4}(M_{V}-M_{P})\label{Eq:rai1-1}
\end{equation}
where $M_{V}$ and $M_{P}$ are the vector and pseudoscalar meson ground state masses. Using this value of $V_{0}$ and $A$, we calculate $S$, $P$ and $D$-state wave spin-averaged masses of $D$ and $D_S$ mesons, as listed in Table (\ref{tab:swavespin}). For the comparison for the $nJ$ state, we compute the spin-average or the center of weight mass from the  respective theoretical values as~\cite{Rai2008}: 

\begin{equation}
M_{CW,n}=\frac{\Sigma_{J}(2J+1)M_{nJ}}{\Sigma_{J}(2J+1)}\label{Eq:rai2-1}
\end{equation}
where $M_{CW,n}$ denotes the spin-averaged mass of the $n$ state and $M_{nJ}$ represents the mass of the meson in the $nJ$ state. 

The value of the QCD coupling constant, potential parameter and the value of the constant $V_{0}$ for $D$ and $D_s$ mesons are given in Table (\ref{tab:potpar}). The quark masses are $m_{u/d}=0.46\; GeV$,  $m_{c}=1.40 \;GeV$, and $m_{s}= 0.586\;GeV$. The quark masses are chosen so as to reproduce the ground state masses of the $D$ and $D_s$ mesons. The value of the charm quark mass $m_c$ chosen in the present work is somewhat lower than that used by most authors in the literature. A higher value of $m_c$  causes the mass spectra to overestimate, but the spectrum can be revised if one uses a confinement potential based on screening instead of the usual linear confining employed here. See for example references \cite{Chao1992,Ding1993,Li:prd2009}.

\subsection{Spin-dependent potential\label{sec:mass2}}
The spin-dependent part of the usual one-gluon exchange potential (OGEP) between the quark and anti-quark for computing the hyperfine and spin-orbit shifting of the low-lying $S$, $P$ and $D$-states is given by \cite{Eichten1994,Gromes1984,Gershtein1995}

\begin{eqnarray}
V_{SD}(\mathbf{r}) & = & \left(\frac{\mathbf{L\cdot S_{Q}}}{2m_{Q}^{2}}+\frac{\mathbf{L\cdot S_{\bar{q}}}}{2m_{\bar{q}}^{2}}\right)\left(-\frac{dV^{\left(0\right)}(r)}{rdr}+\frac{8}{3}\alpha_{S}\frac{1}{r^{3}}\right)+\nonumber \\
 &  & \frac{4}{3}\alpha_{S}\frac{1}{m_{Q}m_{\bar{q}}}\frac{\mathbf{L\cdot S}}{r^{3}}+\frac{4}{3}\alpha_{S}\frac{2}{3m_{Q}m_{\bar{q}}}\mathbf{S_{Q}\cdot S_{\bar{q}}}4\pi\delta(\mathbf{r})\label{eq:spinhyperfine} \nonumber\\
 &  & +\frac{4}{3}\alpha_{S}\frac{1}{m_{Q}m_{\bar{q}}}\Biggl\{3(\mathbf{S_{Q}\cdot n})(\mathbf{S_{\bar{q}}\cdot n})-\nonumber \\
 &  & (\mathbf{S_{Q}\cdot S_{\bar{q}}})\Biggr\}\frac{1}{r^{3}},\ \quad\mathbf{n}=\frac{\mathbf{r}}{r} 
\end{eqnarray}
where $V^{0}(r)$ is the phenomenological potential, the first term accounts for the relativistic corrections to the potential $V^{0}(r)$, the second term accounts for the spin orbital interaction, the third term is the usual spin-spin interaction part which is responsible for pseudoscalar and vector meson splitting, and the fourth term stands for the tensor interaction.

For a meson with unequal quark masses, mass eigenstates are constructed by $jj$ coupling. The angular momentum of the heavy quark is described by its spin $\mathbf{S}_{\mathbf{Q}}$, and that of the light degrees of freedom is described by  \ensuremath{\mathbf{j_{\bar{q}}}=\mathbf{s_{\bar{q}}}+\mathbf{L}}, where \ensuremath{\mathbf{s_{\bar{q}}}}  is the light quark spin and \ensuremath{\mathbf{L}}  is the orbital angular momentum of the light quark. The quantum numbers \ensuremath{\mathbf{S_{Q}}}  and \ensuremath{\mathbf{j_{\bar{q}}}}  are individually conserved. The quantum numbers of the excited \ensuremath{\mathbf{L}=1}  states are formed by combining \ensuremath{\mathbf{S_{Q}}}  and \ensuremath{\mathbf{j_{\bar{q}}}} . For \ensuremath{\mathbf{L}=1} we have \ensuremath{\mathbf{j_{\bar{q}}}=1/2}  \ensuremath{(\mathbf{J}=0,1)}  and \ensuremath{\mathbf{j_{\bar{q}}}=3/2}  \ensuremath{(\mathbf{J}=1,2)}  states. These states are denoted as $^{3}P_{0}$, \ensuremath{^{1}P_{1}^{\prime}}  ( \ensuremath{\mathbf{j_{\bar{q}}}=1/2}  ), \ensuremath{^{1}P_{1}}  \ensuremath{(\mathbf{j_{\bar{q}}}=3/2)}
 and \ensuremath{^{3}P_{2}}  in the case of the $D$ and $D_S$ meson.

Independently of the total spin $J$ projection, one has

\begin{eqnarray}
\left|^{2L+1}L_{L+1}\right\rangle  & = & \left|J=L+1,S=1\right\rangle 
\end{eqnarray}

\begin{eqnarray}
\left|^{2L+1}L_{L}\right\rangle  & = & \sqrt{\frac{L}{L+1}}\left|J=L,S=1\right\rangle +\nonumber \\
 &  & \sqrt{\frac{L+1}{2L+1}}\left|J=L,S=0\right\rangle 
\end{eqnarray}

\begin{eqnarray}
\left|^{2L-1}L_{L}\right\rangle  & = & \sqrt{\frac{L+1}{2L+1}}\left|J=L,S=1\right\rangle -\nonumber \\
 &  & \sqrt{\frac{L}{2L+1}}\left|J=L,S=0\right\rangle 
\end{eqnarray}

\noindent where $\left|J,S\right\rangle $ are the state vectors with the given values of the total quark spin given by \textbf{$\mathbf{S=s_{\bar{q}}+S_{Q}}$, }so that the potential terms of the order of $1/m_{\bar{q}}m_{Q}$, $1/m_{Q}^{2}$, lead to the mixing of the levels with the different $j_{\bar{q}}$ values at the given $J$ values. The tensor forces (the last term in equation (\ref{eq:spinhyperfine}) are equal to zero at $L=0$ or $S=0$.

The heavy-heavy flavored meson states with $J=L$ are mixtures of spin-triplet $\left|^{3}L_{L}\right>$ and spin-singlet $\left|^{1}L_{L}\right>$
states: $J=L=1,\ 2,\ 3,\ldots$

\begin{eqnarray}
\left|\psi_{J}\right> & = & \left|^{1}L_{L}\right>\cos{\phi}+\left|^{3}L_{L}\right>\sin{\phi}
\end{eqnarray}
\begin{eqnarray}
\left|\psi_{J}^{\prime}\right> & = & -\left|^{1}L_{L}\right>\sin{\phi}+\left|^{3}L_{L}\right>\cos{\phi} \end{eqnarray}

\noindent where $\phi$ is the mixing angle and the primed state has the heavier mass. Such mixing occurs due to the nondiagonal spin-orbit and tensor terms in Equation (\ref{eq:spinhyperfine}). The masses of the physical states were obtained by diagonalizing the mixing matrix obtained using equation (\ref{eq:spinhyperfine}) \cite{Gershtein1995}.

\subsection{Leptonic and Radiative Leptonic Branching Fractions\label{sec:lepto}}

The leptonic branching fractions for the ($1^{1}S_{0}$)  mesons are obtained using the formula 

\begin{equation}
BR=\Gamma\times\tau \label{Eq:branchRatio}
\end{equation}

where $\Gamma$ (leptonic decay width) is given by \cite{Silverman1988},

\begin{eqnarray}
\Gamma({D/D_S}^{+}\rightarrow l^{+}\nu_{l})& = & \frac{G_{F}^{2}}{8\pi}f_{D/D_S}^{2}\left|V_{cd/cs}\right|^{2}m_{l}^{2}\times\nonumber \\
& & \left(1-\frac{m_{l}^{2}}{M_{D/D_S}^{2}}\right)^{2}M_{D/D_S}\label{Eq:branchingd}
\end{eqnarray}

For the calculation of the radiative leptonic decay widths of $D^{-}\rightarrow\gamma l\bar{\nu},\;(l=e,\mu)$
 and  $D_{S}\rightarrow\gamma l\bar{\nu},\;(l=e,\mu)$ width, we employ Equations (\ref{Eq:radiativeD}) and (\ref{Eq:radiativeDs})  respectively \cite{Cai-Dian2003}. 

\begin{equation}
\Gamma(D^{-}\rightarrow\gamma l\bar{\nu})=\frac{\alpha G_{F}^{2}\left|V_{cd}\right|^{2}}{2592\pi^{2}}f_{D^{-}}^{2}m_{D^{-}}^{3}\left[x_{d}+x_{c}\right];\label{Eq:radiativeD}
\end{equation}
where
\begin{equation}
x_{d}=\left(3-\frac{m_{D^{-}}}{m_{d}}\right)^{2},\text{ \,\,\,\,\,\,\,\,}x_{c}=\left(3-2\frac{m_{D^{-}}}{m_{c}}\right)^{2}
\end{equation}

\begin{equation}
\Gamma(D_{s}\rightarrow\gamma l\bar{\nu})=\frac{\alpha G_{F}^{2}\left|V_{cs}\right|^{2}}{2592\pi^{2}}f_{D_{s}}^{2}m_{D_{s}}^{3}\left[x_{s}+x_{c}\right];\label{Eq:radiativeDs}
\end{equation}
where
\begin{equation}
x_{s}=\left(3-\frac{m_{D_{S}}}{m_{s}}\right)^{2},\text{ \,\,\,\,\,\,\,\,}x_{c}=\left(3-2\frac{m_{D_{S}}}{m_{c}}\right)^{2}.
\end{equation}

The decay constants for the calculation of leptonic and radiative leptonic branching fractions were obtained from the Van-Royen-Weisskopf formula, incorporating a first order QCD correction factor  \cite{VanRoyen1967}, 

\begin{equation}
f_{P/V}^{2}=\frac{12\left|\psi_{P/V}(0)\right|^{2}}{M_{P/V}}\bar{C^{2}}(\alpha_{S}),\label{Eq:decayconst}
\end{equation}
where \ensuremath{\bar{C^{2}}(\alpha_{S})}
 is the QCD correction factor given by \cite{Braaten1995}:
 
\begin{equation}
\bar{C^{2}}(\alpha_{S})=1-\frac{\alpha_{S}}{\pi}\left[2-\frac{m_{Q}-m_{\bar{q}}}{m_{Q}+m_{\bar{q}}}\ln\frac{m_{Q}}{m_{\bar{q}}}\right].\label{Eq:correction}
\end{equation}

For the calculation of the leptonic and radiative leptonic decay width and corresponding branching fractions using Equation (\ref{Eq:branchRatio}), we take $\tau_{D}=0.410$ ps and $\tau_{D_S}=0.5$ ps \cite{PDGlatest}, and employ the calculated values of the pseudoscalar decay constants (in GeV) $f_{Dcor}=0.165$, $f_{D}=0.247$ for the $D$ meson  and  $f_{D_{Scor}}=0.217$, $f_{D_{S}}=0.324$ for the $D_s$ meson, with and without QCD correction respectively using the masses obtained from Tables ({\ref{tab:massesD}}) and (\ref{tab:massesDs}).

\subsection{Mixing parameters\label{sec:mixing}}
In the Standard Model, the transitions $D_{q}^{0}-\bar{D}_{q}^{0}$ and $\bar{D}_{q}^{0}-D_{q}$ occur due to the weak interaction. The neutral $D$ mesons mix with their antiparticles, leading to oscillations between the mass eigenstates \cite{PDGlatest}. Following notation introduced in Ref.~\cite{PDGlatest} and assuming CPT conservation throughout, in each system, the light (L) and heavy (H) mass eigenstates,

\begin{equation}
\left|D_{q}^{L,H}\right\rangle =\frac{1}{\sqrt{1+\left|\left(q/p\right)_{q}\right|^{2}}}\left(\left|D_{q}\right\rangle \pm\left(q/p\right)_{q}\left|\bar{D}_{q}\right\rangle \right)
\end{equation}

\noindent have a mass difference $\Delta m_{q}=m_{H}-m_{L}>0$, and a total decay width difference $\Delta\Gamma_{q}=\Gamma_{L}-\Gamma_{H}$.
The time evolution of the neutral $D$ meson doublet is described by the Schrodinger equation \cite{HoKim1998gr,Buchalla2008jp}

\begin{equation}
i\frac{d}{dt}\left(\frac{D_{q}}{D_{q}}\right)=\left[\left(\begin{array}{cc}
M_{11}^{q} & M_{12}^{q\star}\\
M_{12}^{q} & M_{11}^{q}
\end{array}\right)-\frac{i}{2}\left(\begin{array}{cc}
\Gamma_{11}^{q} & \Gamma_{12}^{q\star}\\
\Gamma_{12}^{q} & \Gamma_{11}^{q}
\end{array}\right)\right]\left(\frac{D_{q}}{D_{q}}\right)
\end{equation}\\

Here, the two $2\times2$ matrices are a consequence of CPT invariance. The expressions for the off-diagonal elements of the mass and the decay matrices are \cite{Buras1984pq}

\begin{equation}
M_{12}=-\frac{G_{F}^{2}m_{W}^{2}\eta_{D}m_{D_{q}}B_{D_{q}}f_{D_{q}}^{2}}{12\pi^{2}}S_{0}\left(m_{s}^{2}/m_{W}^{2}\right)\left(V_{us}^{\star}V_{cs}\right)^{2}
\end{equation}

\begin{equation}
\Gamma_{12}=\frac{G_{F}^{2}m_{c}^{2}\eta_{D}^{\prime}m_{D_{q}}B_{D_{q}}f_{D_{q}}^{2}}{8\pi}\left[\left(V_{us}^{\star}V_{cs}\right)^{2}\right]
\end{equation}
 where $G_{F}$ is the Fermi constant, $m_{W}$ is the $W$ boson mass, $m_{c}$ is the mass of the $c$ quark, and $m_{D}$, $f_{D_{q}}$ and $B_{D_{q}}$ are the $D^{0}$ mass, the weak decay constant and the bag parameter, respectively. The known function $S_{0}(x_{t})$
can be approximated very well by $0.784x_{t}^{0.76}$ \cite{Inami1980}, and $V_{ij}$ are the elements of the CKM matrix. The parameters $\eta_{D}$ and $\eta_{D}^{\prime}$ correspond to the gluonic corrections.

Theoretically, the hadron lifetime ($\tau_{B}$) is related to $\Gamma_{11}^{q}\left(\tau_{B_{q}}=1/\Gamma_{11}^{q}\right)$, while the observables $\Delta m_{q}$ and $\Delta\Gamma_{q}$ are related to $M_{12}^{q}$ and $\Gamma_{12}^{q}$ as\cite{PDGlatest}

\begin{equation}
\triangle m_{q}=2\left|M_{12}^{q}\right|
\end{equation}
and

\begin{equation}
\triangle\Gamma_{q}=2\left|\Gamma_{12}^{q}\right|
\end{equation}

 The integrated oscillation rate $(\chi_{q})$ is the probability of observing a $\bar{D}$ meson in a jet initiated by a $\bar{c}$ quark, as the mass difference $\Delta m$ is a measure of the frequency of the change from a $D^{0}$ into a $\bar{D}^{0}$ or vice versa.

In the absence of CP violation, we have the time-integrated mixing rate for semi-leptonic decays as

\begin{equation}
R_{M}\simeq\frac{1}{2}(x_{q}^{2}+y_{q}^{2}).
\end{equation}

 \end{multicols}

 \begin{table*}[h]
   \begin{centering}
  \caption{Potential parameters\label{tab:potpar}.}
  \begin{tabular}{cccc}
  \hline 
  Meson & $\alpha_{c}$ & $A\;(GeV^{2})$ & $V_{0}\;(GeV)$\tabularnewline
  \hline 
  $D$ & 0.731 & 0.122 & -0.199\tabularnewline
  \hline 
  $D_{s}$ & 0.633 & 0.137 & -0.169\tabularnewline
  \hline 
  \end{tabular}
  \par\end{centering}
  \end{table*}

 \begin{table*}[h]
    \centering
    \caption{Spin average S-P-D-wave masses of the D and Ds mesons (in GeV).\label{tab:swavespin}}
    
    \begin{tabular}{cccccccccc}
    \hline \noalign{\smallskip}
     
     Meson & State & $\mu$ & $M_{SA}$ & Expt\cite{PDGlatest} & \cite{Godfrey2015} & \cite{Ebert2010} & \cite{Lahde2000} & \cite{Devlani2013, Devlani2011} & \cite{Li2011}\tabularnewline
    \noalign{\smallskip}\hline\noalign{\smallskip}
    
    \multirow{11}{*}{$D$} & $1S$ & 0.390 & 1.975 & 1.975 & 2.000 & 1.975 & 1.973 & 1.975 & 1.974\tabularnewline
     & $2S$ & 0.281 & 2.636 & 2.613 & 2.628 & 2.619 & 2.586 & 2.586 & 2.616\tabularnewline
     & $3S$ & 0.233 & 3.225 &  & 3.100 & 3.087 & 2.936 & 3.104 & \tabularnewline
     & $4S$ & 0.206 & 3.778 &  & 3.490 & 3.474 & 3.208 & 3.510 & \tabularnewline
     & $5S$ & 0.187 & 4.310 &  & 3.820 & 3.815 &  &  & \tabularnewline
     \noalign{\smallskip}
     
     & $1P$ & 0.309 & 2.440 & 2.434 & 2.473 & 2.414 & 2.426 & 2.448 & 2.420\tabularnewline
     & $2P$ & 0.247 & 3.027 &  & 2.948 & 2.986 & 3.020 & 2.949 & 2.920\tabularnewline
     & $3P$ & 0.215 & 3.575 &  & 3.348 & 3.405 & 3.105 &  & \tabularnewline
     \noalign{\smallskip}
    
     & $1D$ & 0.276 & 2.779 &  & 2.830 & 2.834 & 2.707 & 2.768 & 2.733\tabularnewline
     & $2D$ & 0.231 & 3.338 &  & 3.229 & 3.293 & 3.014 & 3.207 & 3.175\tabularnewline
     & $3D$ & 0.204 & 3.870 &  & 3.582 &  &  &  & \tabularnewline
     
    \hline\noalign{\smallskip}
    
    \multirow{11}{*}{$Ds$} & $1S$ & 0.477 & 2.076 & 2.076 & 2.091 & 2.075 & 2.074 & 2.076 & 2.072\tabularnewline
     & $2S$ & 0.339 & 2.709 &  & 2.717 & 2.720 & 2.706 & 2.713 & 2.695\tabularnewline
     & $3S$ & 0.282 & 3.261 &  & 3.222 & 3.236 & 3.076 & 3.175 & \tabularnewline
     & $4S$ & 0.249 & 3.772 &  & 3.568 & 3.665 & 3.356 & 3.567 & \tabularnewline
     & $5S$ & 0.227 & 4.260 &  & 3.898 & 4.044 &  &  & \tabularnewline
     \noalign{\smallskip}
     
     & $1P$ & 0.368 & 2.542 & 2.514 & 2.563 & 2.537 & 2.538 & 2.540 & 2.511\tabularnewline
     & $2P$ & 0.296 & 3.090 &  & 3.034 & 3.119 & 2.954 & 3.026 & 2.991\tabularnewline
     & $3P$ & 0.259 & 3.595 &  & 3.430 & 3.569 & 3.255 &  & \tabularnewline
     \noalign{\smallskip}
     
     & $1D$ & 0.326 & 2.871 &  & 2.912 & 2.950 & 2.850 & 2.852 & 2.814\tabularnewline
     & $2D$ & 0.276 & 3.383 &  & 3.310 & 3.436 & 3.161 & 3.277 & 3.236\tabularnewline
     & $3D$ & 0.246 & 3.869 &  & 3.661 &  &  &  & \tabularnewline
    
    \hline 
    \multicolumn{10}{l}{Ref.~\cite{Devlani2013} for the $D$ meson and Ref.~{\cite{Devlani2011}} for the $D_S$ meson}\tabularnewline
  
    \end{tabular}
    \end{table*}

 \begin{multicols}{2}

 \section{Results and Discussion\label{sec:Resu}}
 
 The spin averaged masses for the S, P and D states are tabulated in Table~(\ref{tab:swavespin}), and are in good agreement with experimental as well as other theoretical model predictions. The calculated values of the mass spectra of the $D$ and $D_S$ mesons are listed in Tables (\ref{tab:massesD} and \ref{tab:massesDs}). We follow the spectroscopic notation $n^{2S+1}L_{J}$ in the tables. The results obtained using the present framework are improved in comparison to the previous calculations outlined in Ref.~\cite{Devlani2013,Devlani2011}. The mass spectra of charmed and charmed-strange mesons are also shown graphically in Figs.~(\ref{fig:MassD}) and (\ref{fig:MassDs}). The obtained results for charmed mesons are close to the experimental measurements, especially for excited states. The charmed-strange meson spectra are also close to the experimental measurements, but the $1^{3}P_{0}$ state for the $D_s$ meson is overestimated by about 120 MeV.

 The calculated results of leptonic branching fractions and radiative leptonic decay widths for the $D$ and $D_S$ mesons are tabulated in Tables~(\ref{tab:leptobranch}) and (\ref{tab:radiativelepto}) respectively. The calculated results of leptonic branching fractions show that the predictions are fairly close to the experimental results. The obtained results of leptonic branching fractions  for the $D_s$ meson  are slightly underestimated with respect to experimental observations.
   
 In the literature, various methods are used to calculate the radiative leptonic decay rates and branching ratios. In Ref.~\cite{Atwood1996}, 
 $Ds\rightarrow l\bar{\nu}{\gamma}$ is calculated in a non-relativistic quark model, and the branching ratio obtained is of the order $10^{-4}$. In Ref.~\cite{Korchemsky:2000}, with the perturbative QCD approach, it is found that the branching ratio of $D_{s}^{+}\rightarrow e^{+}\bar{\nu}{\gamma}$
 is of the order of $10^{-3}$ and
 $D^{+}\rightarrow e^{+}\bar{\nu}{\gamma}$
 of the order of $10^{-4}$. In Ref.~\cite{Geng:2000}, the
light front quark model estimate is of the order of $10^{-6}$ and in the non-relativistic constituent quark model, the branching ratio of
 $D^{-}\rightarrow l\bar{\nu}{\gamma}$ is of the order of $10^{-6}$ and
 $D_{s}^{-}\rightarrow l\bar{\nu}{\gamma}$  of the order of $10^{-5}$ \cite{Cai-Dian2003}. In the factorization approach it is found to be of the order $10^{-5}$  for D meson \cite{Yang:2012,Yang:2014,Yang:2015,Yang:2016}. In the present work the branching ratios are of the order of $10^{-6}$ in both the cases. All the various approaches give different branching ratios in the range of the order $10^{-3}$ to $10^{-6}$.
  
For the estimation of the mixing parameters $x_{q}$, $y_{q}$, $\chi_{q}$ and $R_{M}$, we use $\eta_{D}=0.86$, $\eta_{D}^{\prime}=0.21$, and the gluonic correction to the oscillation is given by Ref.~\cite{Shah2016,Buras1990fn}. The bag parameter $B_{D_{q}}=1.34$ is taken from the lattice result of \cite{Buras2003td}, while the pseudoscalar mass ($M_{D_{q}}$) and the pseudoscalar decay constant ($f_{D_{q}}$) of the charmed mesons are taken from our present study. The values of $m_{s}$ (0.1~GeV), $m_{W}$ (80.399~GeV), the experimental average lifetime of  the $D$ meson, and the CKM matrix elements $V_{us}$ (0.22522) and $V_{cs}$ (0.97427), are taken from the Particle Data Group \cite{PDGlatest}. The calculated  mixing parameters $x_{q}$, $y_{q}$, $\chi_{q}$ and $R_{M}$ are tabulated in Table~(\ref{tab:mixing}), along with the experimental results. Our results are in accordance with the BaBar(2010) experimental results.  

\end{multicols}

  \begin{center}
   \includegraphics[bb=30bp 60bp 750bp 550bp,clip,width=0.72
   \textwidth]{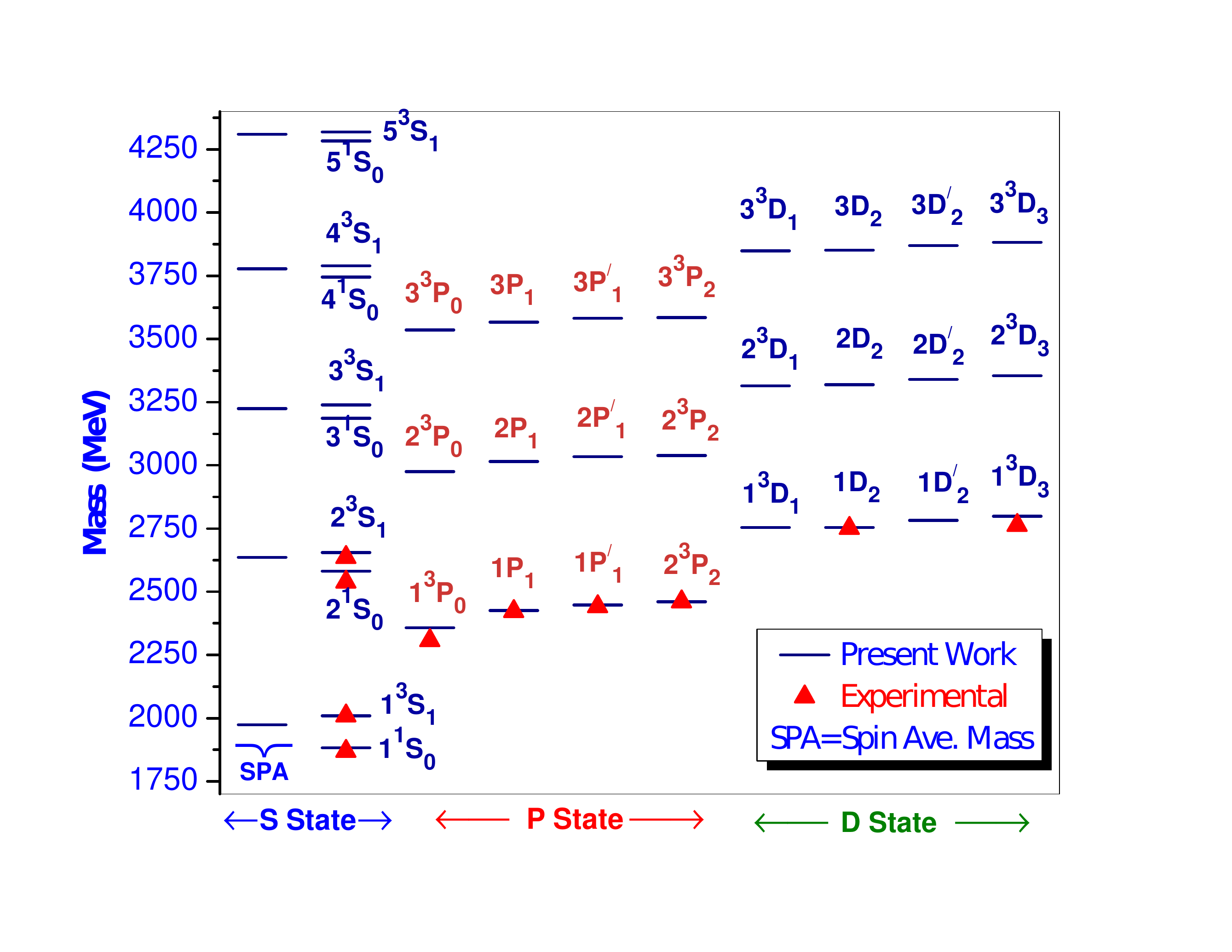}
   \figcaption{\label{fig:MassD}   Mass spectrum of charmed mesons.}
  \end{center}
   
  \begin{center}
  \includegraphics[bb=30bp 60bp 750bp 550bp,clip,width=0.72
   \textwidth]{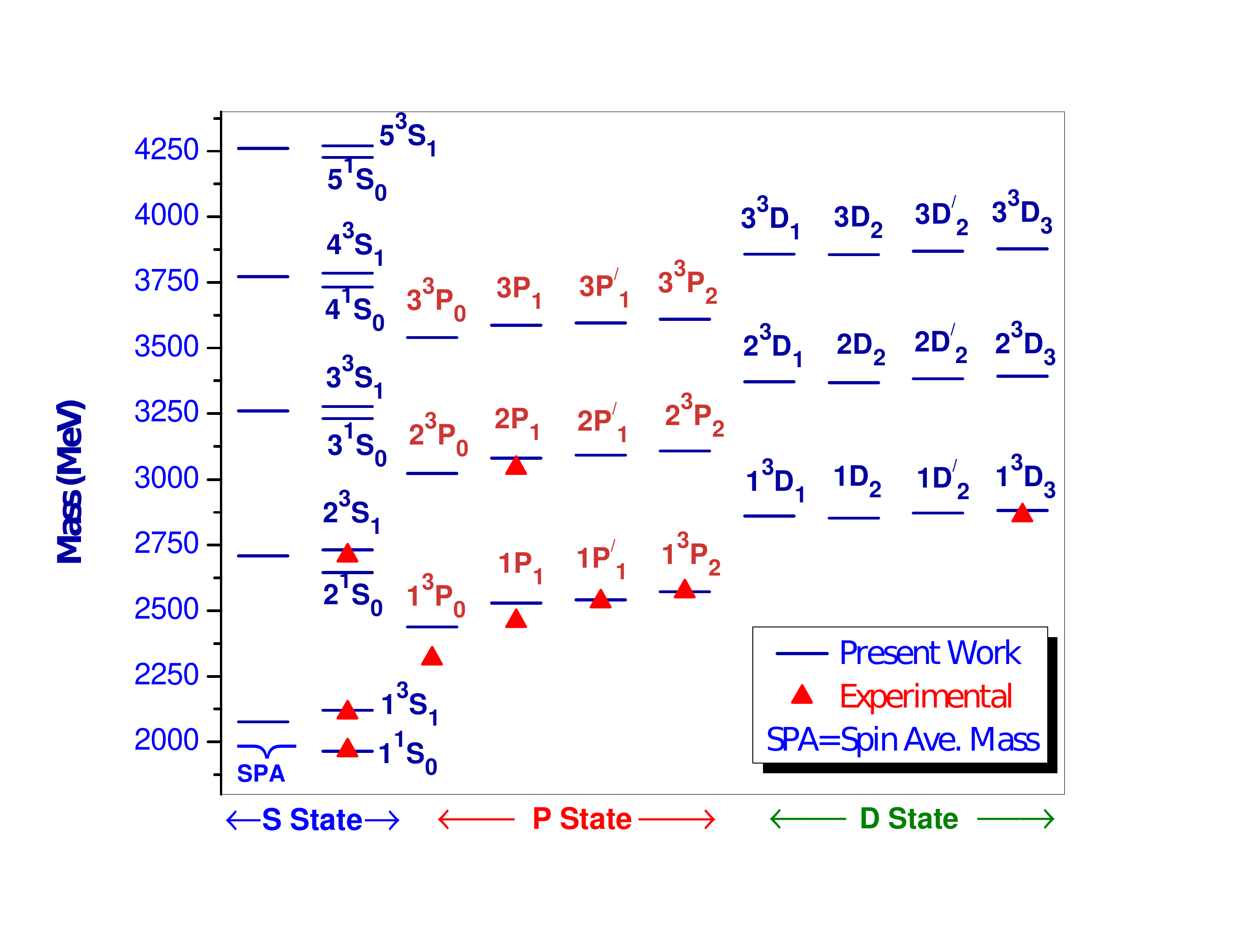}
  \figcaption{\label{fig:MassDs}   Mass spectrum of charmed-strange mesons. }
  \end{center}

  \noindent 
  \begin{table*}
  \centering
  \caption{Mass spectrum of the $D$ mesons (in GeV).\label{tab:massesD}}
  
  \centering{}%
  \begin{tabular}{lllllllllll}
  \hline \noalign{\smallskip}
  State & \multirow{2}{*}{$J^{P}$} & \multicolumn{1}{l}{Present} & \multirow{2}{*}{Expt.\cite{PDGlatest}} & \multirow{2}{*}{LATTICE\cite{Cichy2016}} & \multirow{2}{*}{\cite{Godfrey2015}} & \multirow{2}{*}{\cite{Devlani2013}} & \multirow{2}{*}{\cite{Ebert2010}} & \multirow{2}{*}{\cite{Lahde2000}} & \multirow{2}{*}{\cite{Pierro2001}} & \multirow{2}{*}{\cite{Li2011}}\tabularnewline
  $n^{2S+1}L_{J}$ &  &  work &  &  &  &  &  &  &  & \tabularnewline
  \noalign{\smallskip}\hline\noalign{\smallskip}
  
  $1^{1}S_{0}$ & $0^{-}$ & 1.884 & 1.864$(D^{0})$ & 1.865 & 1.877 & 1.865 & 1.871 & 1.874 & 1.868 & 1.867\tabularnewline
   &  &  & 1.870$(D^{\pm})$ &  &  &  &  &  &  & \tabularnewline
  $1^{3}S_{1}$ & $1^{-}$ & 2.010 & 2.007$(D^{\star}(2007)^{0})$ & 2.027 & 2.041 & 2.018 & 2.010 & 2.006 & 2.005 & 2.010\tabularnewline
   &  &  & 2.010$(D^{\star}(2010)^{\pm})$ &  &  &  &  &  &  & \tabularnewline
  $2^{1}S_{0}$ & $0^{-}$ & 2.582 & 2.539 &  & 2.581 & 2.598 & 2.581 & 2.540 & 2.589 & 2.555\tabularnewline
  $2^{3}S_{1}$ & $1^{-}$ & 2.655 & 2.637 &  & 2.643 & 2.639 & 2.632 & 2.601 & 2.692 & 2.636\tabularnewline
  $3^{1}S_{0}$ & $0^{-}$ & 3.186 &  &  & 3.068 & 3.087 & 3.062 & 2.904 & 3.141 & \tabularnewline
  $3^{3}S_{1}$ & $1^{-}$ & 3.239 &  &  & 3.110 & 3.110 & 3.096 & 2.947 & 3.226 & \tabularnewline
  $4^{1}S_{0}$ & $0^{-}$ & 3.746 &  &  & 3.468 & 3.498 & 3.452 & 3.175 &  & \tabularnewline
  $4^{3}S_{1}$ & $1^{-}$ & 3.789 &  &  & 3.497 & 3.514 & 3.482 & 3.208 &  & \tabularnewline
  $5^{1}S_{0}$ & $0^{-}$ & 4.283 &  &  & 3.814 &  & 3.793 &  &  & \tabularnewline
  $5^{3}S_{1}$ & $1^{-}$ & 4.319 &  &  & 3.837 &  & 3.822 &  &  & \tabularnewline
  
  \hline\noalign{\smallskip}
  
  $1^{3}P_{0}$ & $0^{+}$ & 2.357 & $2.318\pm0.029$ & 2.325 & 2.399 & 2.352 & 2.406 & 2.341 & 2.377 & 2.252\tabularnewline
  $1P_{1}$ & $1^{+}$ & 2.425 & 2.421 & 2.468 & 2.456 & 2.434 & 2.426 & 2.389 & 2.417 & 2.402\tabularnewline
  $1P_{1}^{\prime}$ & $1^{+}$ & 2.447  & $2.441\pm0.032$ & 2.631 & 2.467 & 2.454 & 2.469 & 2.407 & 2.490 & 2.417\tabularnewline
  $1^{3}P_{2}$ & $2^{+}$ & 2.461 & 2.463 & 2.743 & 2.502 & 2.473 & 2.460 & 2.477 & 2.460 & 2.466\tabularnewline
  $2^{3}P_{0}$ & $0^{+}$ & 2.976 &  &  & 2.931 & 2.868 & 2.919 & 2.758 & 2.949 & 2.752\tabularnewline
  $2P_{1}$ & $1^{+}$ & 3.016 &  &  & 2.924 & 2.940 & 2.932 & 2.792 & 2.995 & 2.886\tabularnewline
  $2P_{1}^{\prime}$ & $1^{+}$ & 3.034 &  &  & 2.961 & 2.951 & 3.021 & 2.802 & 3.045 & 2.926\tabularnewline
  $2^{3}P_{2}$ & $2^{+}$ & 3.039 &  &  & 2.957 & 2.971 & 3.012 & 2.860 & 3.035 & 2.971\tabularnewline
  $3^{3}P_{0}$ & $0^{+}$ & 3.536 &  &  & 3.343 &  & 3.346 & 3.050 &  & \tabularnewline
  $3P_{1}$ & $1^{+}$ & 3.567 &  &  & 3.360 &  & 3.461 & 3.082 &  & \tabularnewline
  $3P_{1}^{\prime}$ & $1^{+}$ & 3.582 &  &  & 3.328 &  & 3.365 & 3.085 &  & \tabularnewline
  $3^{3}P_{2}$ & $2^{+}$ & 3.584 &  &  & 3.353 &  & 3.407 & 3.142 &  & \tabularnewline
  
  \hline\noalign{\smallskip}
  
  $1^{3}D_{1}$ & $1^{-}$ & 2.755 &  &  & 2.833 & 2.803 & 2.788 & 2.750 & 2.795 & 2.740\tabularnewline
  $1D_{2}$ & $2^{-}$ & 2.754 & 2.752 &  & 2.816 & 2.722 & 2.806 & 2.689 & 2.775 & 2.693\tabularnewline
  $1D_{2}^{\prime}$ & $2^{-}$ & 2.783 &  &  & 2.845 & 2.829 & 2.850 & 2.727 & 2.883 & 2.789\tabularnewline
  $1^{3}D_{3}$ & $3^{-}$ & 2.788 & 2.761 &  & 2.833 & 2.741 & 2.863 & 2.688 & 2.799 & 2.719\tabularnewline
  $2^{3}D_{1}$ & $1^{-}$ & 3.315 &  &  & 3.231 & 3.233 & 3.228 & 3.052 &  & 3.168\tabularnewline
  $2D_{2}$ & $2^{-}$ & 3.318 &  &  & 3.212 & 3.169 & 3.259 & 2.997 &  & 3.145\tabularnewline
  $2D_{2}^{\prime}$ & $2^{-}$ & 3.341 &  &  & 3.248 & 3.256 & 3.307 & 3.029 &  & 3.215\tabularnewline
  $2^{3}D_{3}$ & $3^{-}$ & 3.355 &  &  & 3.226 & 3.187 & 3.335 & 2.999 &  & 3.170\tabularnewline
  $3^{3}D_{1}$ & $1^{-}$ & 3.850 &  &  & 3.579 &  &  &  &  & \tabularnewline
  $3D_{2}$ & $2^{-}$ & 3.854 &  &  & 3.566 &  &  &  &  & \tabularnewline
  $3D_{2}^{\prime}$ & $2^{-}$ & 3.873 &  &  & 3.600 &  &  &  &  & \tabularnewline
  $3^{3}D_{3}$ & $3^{-}$ & 3.885 &  &  & 3.588 &  &  &  &  & \tabularnewline
  \hline 
  \end{tabular}
  \end{table*}

  \begin{table*}
  \centering
  \caption{Mass spectrum of the $D_{S}$ mesons (in GeV).\label{tab:massesDs}}
  
  \centering{}%
  \begin{tabular}{lllllllllll}
  \hline \noalign{\smallskip}
  State & \multirow{2}{*}{$J^{P}$} & Present & \multirow{2}{*}{Expt.\cite{PDGlatest}} & \multirow{2}{*}{LATTICE\cite{Cichy2016}} & \multirow{2}{*}{\cite{Godfrey2015}} & \multirow{2}{*}{\cite{Devlani2011}} & \multirow{2}{*}{\cite{Ebert2010}} & \multirow{2}{*}{\cite{Lahde2000}} & \multirow{2}{*}{\cite{Pierro2001}} & \multirow{2}{*}{\cite{Li2011}}\tabularnewline
  $n^{2S+1}L_{J}$ &  & work &  &  &  &  &  &  &  & \tabularnewline
  \hline\noalign{\smallskip} 
  
  $1^{1}S_{0}$ & $0^{-}$ & 1.965 & 1.968$(D_{S}^{\pm})$ & 1.968 & 1.979 & 1.970 & l.969 & l.975 & 1.965 & 1.969\tabularnewline
  
  $1^{3}S_{1}$ & $1^{-}$ & 2.120 & 2.112 $(D_{S}^{\star\pm})$ & 2.123 & 2.129 & 2.117 & 2.111 & 2.108 & 2.113& 2.107\tabularnewline
  
  $2^{1}S_{0}$ & $0^{-}$ & 2.680 &  &  & 2.673 & 2.684 & 2.688 & 2.659 & 2.700 & 2.640\tabularnewline
  $2^{3}S_{1}$ & $1^{-}$ & 2.719 & $2.710_{-0.007}^{+0.012}$ &  & 2.732 & 2.723 & 2.731 & 2.722 & 2.806 & 2.714\tabularnewline
  $3^{1}S_{0}$ & $0^{-}$ & 3.247 &  &  & 3.154 & 3.158 & 3.219 & 3.044 & 3.259 & \tabularnewline
  $3^{3}S_{1}$ & $1^{-}$ & 3.265 &  &  & 3.193 & 3.180 & 3.242 & 3.087 & 3.345 & \tabularnewline
  $4^{1}S_{0}$ & $0^{-}$ & 3.764  &  &  & 3.547 & 3.556 & 3.652 & 3.331 &  & \tabularnewline
  $4^{3}S_{1}$ & $1^{-}$ & 3.775  &  &  & 3.575 & 3.571 & 3.669 & 3.364 &  & \tabularnewline
  $5^{1}S_{0}$ & $0^{-}$ & 4.280 &  &  & 3.894 &  & 4.033 &  &  & \tabularnewline
  $5^{3}S_{1}$ & $1^{-}$ & 4.318 &  &  & 3.912 &  & 4.048 &  &  & \tabularnewline
  
  \hline \noalign{\smallskip}
  
  $1^{3}P_{0}$ & $0^{+}$ & 2.438  & 2.318 & 2.390 & 2.484 & 2.444 & 2.509 & 2.455 & 2.487 & 2.344\tabularnewline
  $1P_{1}$ & $1^{+}$ & 2.529  & 2.460 & 2.556 & 2.549 & 2.530  & 2.536 & 2.502 & 2.535 & 2.488\tabularnewline
  $1P_{1}^{\prime}$ & $1^{+}$ & 2.541  & 2.535 & 2.617 & 2.556 & 2.540  & 2.574 & 2.522 & 2.605 & 2.510\tabularnewline
  $1^{3}P_{2}$ & $2^{+}$ & 2.569  & 2.573 & 2.734 & 2.592 & 2.566 & 2.571 & 2.586 & 2.581 & 2.559\tabularnewline
  $2^{3}P_{0}$ & $0^{+}$ & 3.022  &  &  & 3.005 & 2.947  & 3.054 & 2.901 & 3.067 & 2.830\tabularnewline
  $2P_{1}$ & $1^{+}$ & 3.081  & $3.044_{-0.009}^{+0.030}$ &  & 3.018 & 3.019  & 3.067 & 2.928 & 3.114 & 2.958\tabularnewline
  $2P_{1}^{\prime}$ & $1^{+}$ & 3.092  &  &  & 3.038 & 3.023  & 3.154 & 2.942 & 3.165 & 2.995\tabularnewline
  $2^{3}P_{2}$ & $2^{+}$ & 3.109  &  &  & 3.048 & 3.048  & 3.142 & 2.98 & 3.157 & 3.040\tabularnewline
  $3^{3}P_{0}$ & $0^{+}$ & 3.541 &  &  & 3.412 &  & 3.513 & 3.214 &  & \tabularnewline
  $3P_{1}$ & $1^{+}$ & 3.587 &  &  & 3.416 &  & 3.519 & 3.234 &  & \tabularnewline
  $3P_{1}^{\prime}$ & $1^{+}$ & 3.596 &  &  & 3.433 &  & 3.618 & 3.244 &  & \tabularnewline
  $3^{3}P_{2}$ & $2^{+}$ & 3.609 &  &  & 3.439 &  & 3.580 & 3.283 &  & \tabularnewline
  
  \hline \noalign{\smallskip}
  
  $1^{3}D_{1}$ & $1^{-}$ & 2.882 &  &  & 2.899 & 2.873 & 2.9l3 & 2.838 & 2.900 & 2.804\tabularnewline
  $1D_{2}$ & $2^{-}$ & 2.853  &  &  & 2.900 & 2.816  & 2.931 & 2.845 & 2.9l3 & 2.788\tabularnewline
  $1D_{2}^{\prime}$ & $2^{-}$ & 2.872  &  &  & 2.926 & 2.896  & 2.961 & 2.856 & 2.953 & 2.849\tabularnewline
  $1^{3}D_{3}$ & $3^{-}$ & 2.860  & $2.863_{-0.0026}^{+0.004}$ &  & 2.917 & 2.834  & 2.971 & 2.857 & 2.925 & 2.811\tabularnewline
  $2^{3}D_{1}$ & $1^{-}$ & 3.394  &  &  & 3.306 & 3.292 & 3.383 & 3.144 &  & 3.217\tabularnewline
  $2D_{2}$ & $2^{-}$ & 3.368  &  &  & 3.298 & 3.312 & 3.456 & 3.167 &  & 3.217\tabularnewline
  $2D_{2}^{\prime}$ & $2^{-}$ & 3.384  &  &  & 3.323 & 3.248 & 3.403 & 3.172 &  & 3.260\tabularnewline
  $2^{3}D_{3}$ & $3^{-}$ & 3.372  &  &  & 3.311 & 3.263 & 3.469 & 3.157 &  & 3.240\tabularnewline
  $3^{3}D_{1}$ & $1^{-}$ & 3.858 &  &  & 3.661 &  &  &  &  & \tabularnewline
  $3D_{2}$ & $2^{-}$ & 3.857 &  &  & 3.650 &  &  &  &  & \tabularnewline
  $3D_{2}^{\prime}$ & $2^{-}$ & 3.869 &  &  & 3.672 &  &  &  &  & \tabularnewline
  $3^{3}D_{3}$ & $3^{-}$ & 3.878 &  &  & 3.658 &  &  &  &  & \tabularnewline
  \hline 
  \end{tabular}
  \end{table*}

 \begin{table*}[h]
  \caption{Leptonic branching fractions\label{tab:leptobranch}.}
  \centering{}%
 \begin{tabular}{ccccccc}
 \hline  \noalign{\smallskip}
   &  & $D$ &  &  & $D_{S}$ & \tabularnewline
   & $D^{+}\rightarrow\tau^{+}\nu_{\tau}$ & $D^{+}\rightarrow\mu^{+}\nu_{\mu}$ & $D^{+}\rightarrow e^{+}\nu_{e}$ & $D_{S}^{+}\rightarrow\tau^{+}\nu_{\tau}$ & $D_{S}^{+}\rightarrow\mu^{+}\nu_{\mu}$ & $D_{S}^{+}\rightarrow e^{+}\nu_{e}$\tabularnewline
   & $BR_{\tau}\times10^{-3}$ & $BR_{\mu}\times10^{-4}$ &  $ {\bf BR_{e}}$  & $BR_{\tau}\times10^{-2}$ & $BR_{\mu}\times10^{-3}$ & ${\bf BR_{e} }$\tabularnewline
   \noalign{\smallskip}\hline 
   \noalign{\smallskip}
   This Work & 0.86 & 2.47 &  ${\bf 0.58\times10^{-8}}$  & 3.78 & 4.00 & ${\bf 0.94\times10^{-7}}$ \tabularnewline
   \noalign{\smallskip}
   \noalign{\smallskip}
   PDG\cite{PDGlatest} & $<1.2$ & $3.74\pm0.17$ & $<8.8\times10^{-6}$ & $5.55\pm0.24$ & $5.56\pm0.25$ & $<8.3\times10^{-5}$\tabularnewline
    \noalign{\smallskip}
   \hline 
 
   \end{tabular}
   \end{table*}

   \begin{table}
    \begin{centering}
    \caption{Radiative leptonic decay widths and branching ratio.\label{tab:radiativelepto}}
    \begin{tabular}{ccc}
    \hline  \noalign{\smallskip}
    \noalign{\smallskip} Meson & $\Gamma$$(GeV)$ & $BR$\tabularnewline
      \noalign{\smallskip}
    \hline  \noalign{\smallskip}
    $D$  & $0.46\times10^{-18}$ & $0.73\times10^{-6}$\tabularnewline
     \noalign{\smallskip}
      \noalign{\smallskip}
    $D_{s}$  & $2.13\times10^{-18}$ & $1.62\times10^{-6}$\tabularnewline
     \noalign{\smallskip}
    \hline 
   
    \end{tabular}
    \par\end{centering}
    \end{table} 
         
 \begin{table*}
 \caption{Mixing parameters of the $D$ meson.\label{tab:mixing}}
 \centering{}%
 \begin{tabular}{cccccc}
  \hline 
 \noalign{\smallskip}
  & $\Delta m_{q}\times10^{-15}$ & $x_{q}\times10^{-3}$ & $y_{q}\times10^{-3}$ & $\chi_{q}\times10^{-5}$ & $R_{M}\times10^{-3}$\tabularnewline
\noalign{\smallskip} 
 \cline{2-6}\noalign{\smallskip}\noalign{\smallskip}
 This work & $1.27$ & $1.322$ & $7.67$ & $3.03$ & $0.03027$\tabularnewline
  \noalign{\smallskip}
 \noalign{\smallskip}
 BaBar(2010)  & & $1.6\pm2.3\pm1.2\pm0.8$ \cite{AmoSanchez2010} & $5.7\pm2.0\pm1.3\pm0.7$ \cite{AmoSanchez2010} &  & $0.0864\pm0.0311$\cite{Zhang2007}\tabularnewline
  Belle(2007) & & $8.0\pm2.9$\cite{Zhang2007} & $3.3\pm2.4$ \cite{Zhang2007} &  & $0.13\pm0.22\pm0.20$
  \cite{Bitenc2008} \tabularnewline
 CLEO(2005) & & $4.2\pm2.1$ \cite{Hfag} & $4.6\pm1.9$ \cite{Hfag} &  & $0.04_{-0.6}^{+0.7}$ \cite{Aubert2007} \tabularnewline
  & &  &  &  & $0.02\pm0.47\pm0.14$ \cite{Bitenc2005} \tabularnewline
  & &  &  &  & $1.6\pm2.9\pm2.9$ \cite{Cawlfield:2005} \tabularnewline
  \hline 
 \end{tabular}
 \end{table*}

  \begin{multicols}{2}

\subsection{Regge trajectories \label{sec:reg}}
    
    We construct the Regge trajectories in the $(J,M^{2})$ and $(n_{r},M^{2})$ planes using our calculated masses for both orbitally and radially excited heavy-light mesons. We use the following definitions  \cite{Ebert2010}; 
    
    a) The $(J,M^{2})$ Regge trajectory:
    \begin{equation}
     	J=\alpha M^{2}+\alpha_{0}\label{eq:J regge}
    \end{equation}
     
    b) The $(n_{r},M^{2})$ Regge trajectory:
    \begin{equation}
    	n_{r}\equiv n-1=\beta M^{2}+\beta_{0}\label{eq:nr regge};   \end{equation} where  $\alpha,$ $\beta,$ are the slopes and $\alpha_{0},$ $\beta_{0},$ are the intercepts.  
    
 We plot the Regge trajectories in the $(J,M^{2})$ plane for mesons with natural $(P=(-1)^{J})$ and unnatural $(P=(-1)^{J-1})$ parity as shown in Figs.~(\ref{fig:NPmesonD}-\ref{fig:UNPmesonDs}). The Regge trajectories in the $(n_{r},M^{2})$ plane are presented in Figs.~(\ref{fig:PsVmesonD}-\ref{fig:SavmesonDs}). The calculated masses   are shown by hollow squares and the available experimental data are given by dots  with the corresponding meson names. Straight lines were obtained by $\chi^{2}$ fit of the calculated values.
    
 The fitted slopes and intercepts of the Regge trajectories are given in Tables (\ref{tab:alfa},\ref{tab:bita}). From comparison of the slopes in Tables (\ref{tab:alfa},\ref{tab:bita},\ref{tab:Spinave}), we see that the slope values of $\alpha$ are larger then the slope values of $\beta$. The ratio of the mean of $\alpha$ and  $\beta$ is 1.69 and 1.57 for the $D$ and $D_S$ mesons respectively. We see that the calculated heavy-light meson masses fit nicely to the linear trajectories in both planes and are almost parallel to and equidistant with each other.
  
   \end{multicols}

     \begin{center}
     \includegraphics[bb=30bp 40bp 750bp 575bp,clip,width=0.72
     \textwidth]{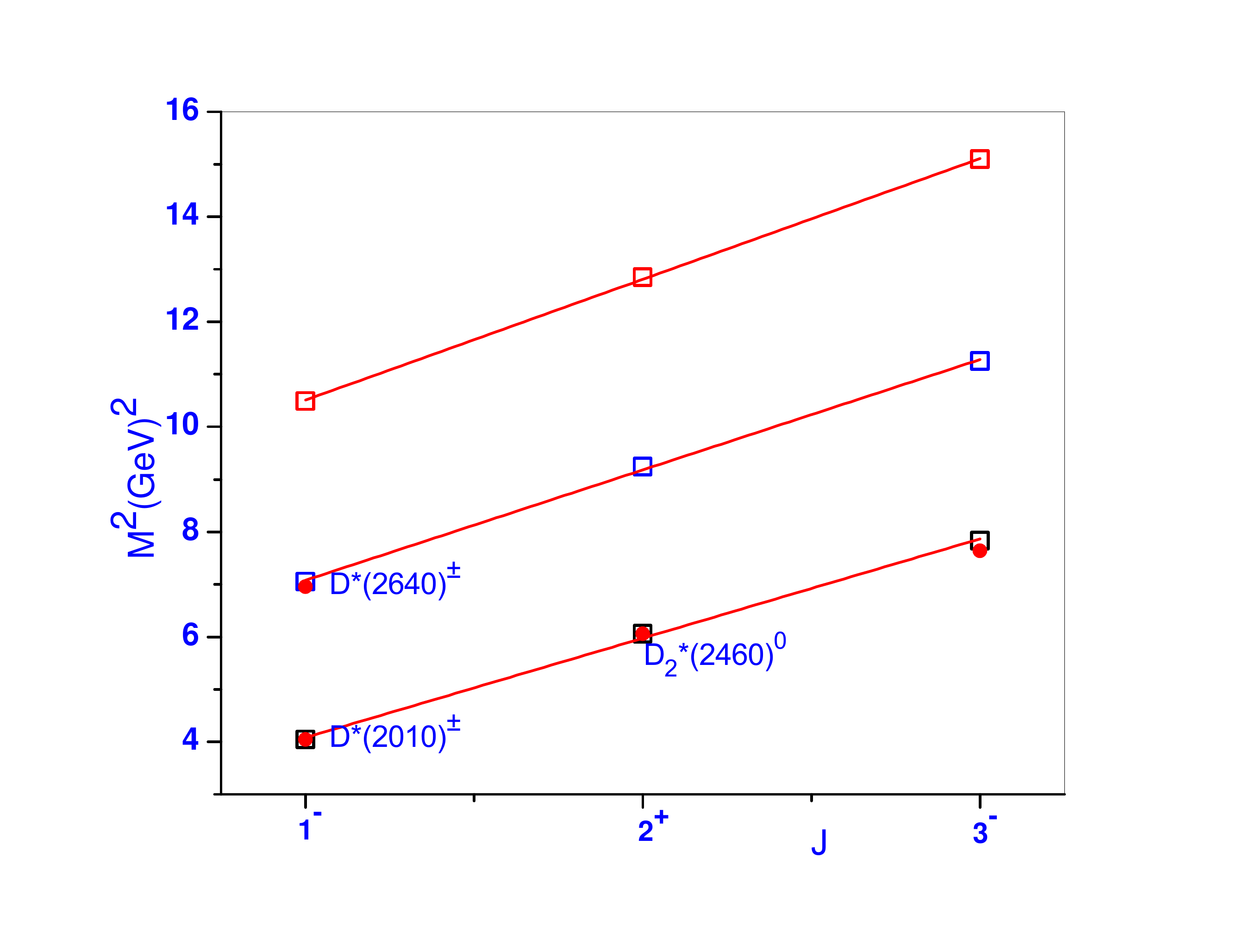}
     \figcaption{\label{fig:NPmesonD}   Parent and daughter $(J,\: M^{2})$ Regge trajectories for charmed mesons with natural parity. Hollow squares are predicted masses. Available experimental data are given by dots with particle names. }
     \end{center}

     \begin{center}
     \includegraphics[bb=30bp 40bp 750bp 575bp,clip,width=0.72
     \textwidth]{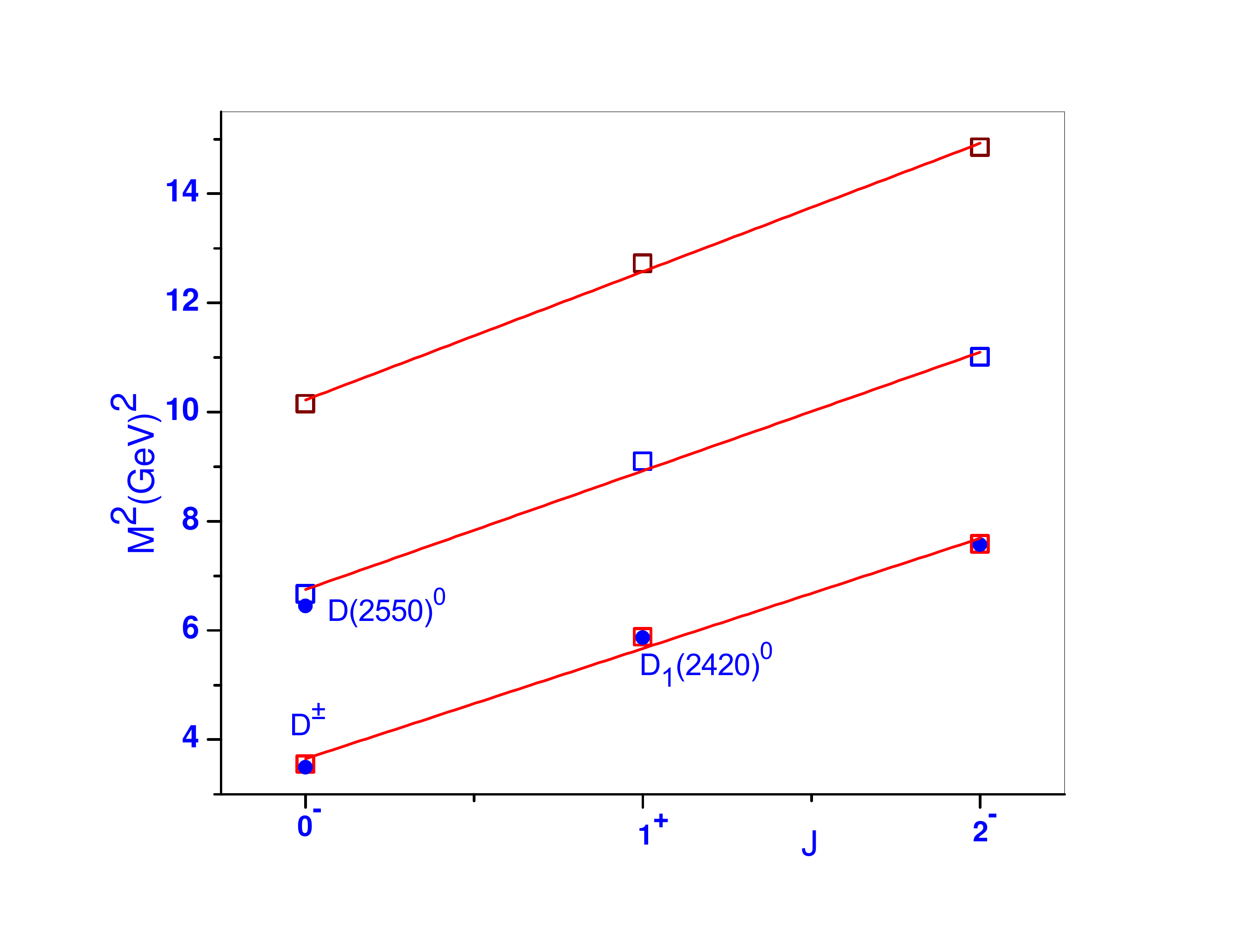}
     \figcaption{\label{fig:UNPmesonD} Parent and daughter $(J,\: M^{2})$ Regge trajectories for  charmed mesons with unnatural parity. Hollow squares are predicted masses. Available experimental data are given by dots with particle names.}
      \end{center}

   \begin{center}
    \includegraphics[bb=30bp 40bp 750bp 575bp,clip,width=0.72
    \textwidth]{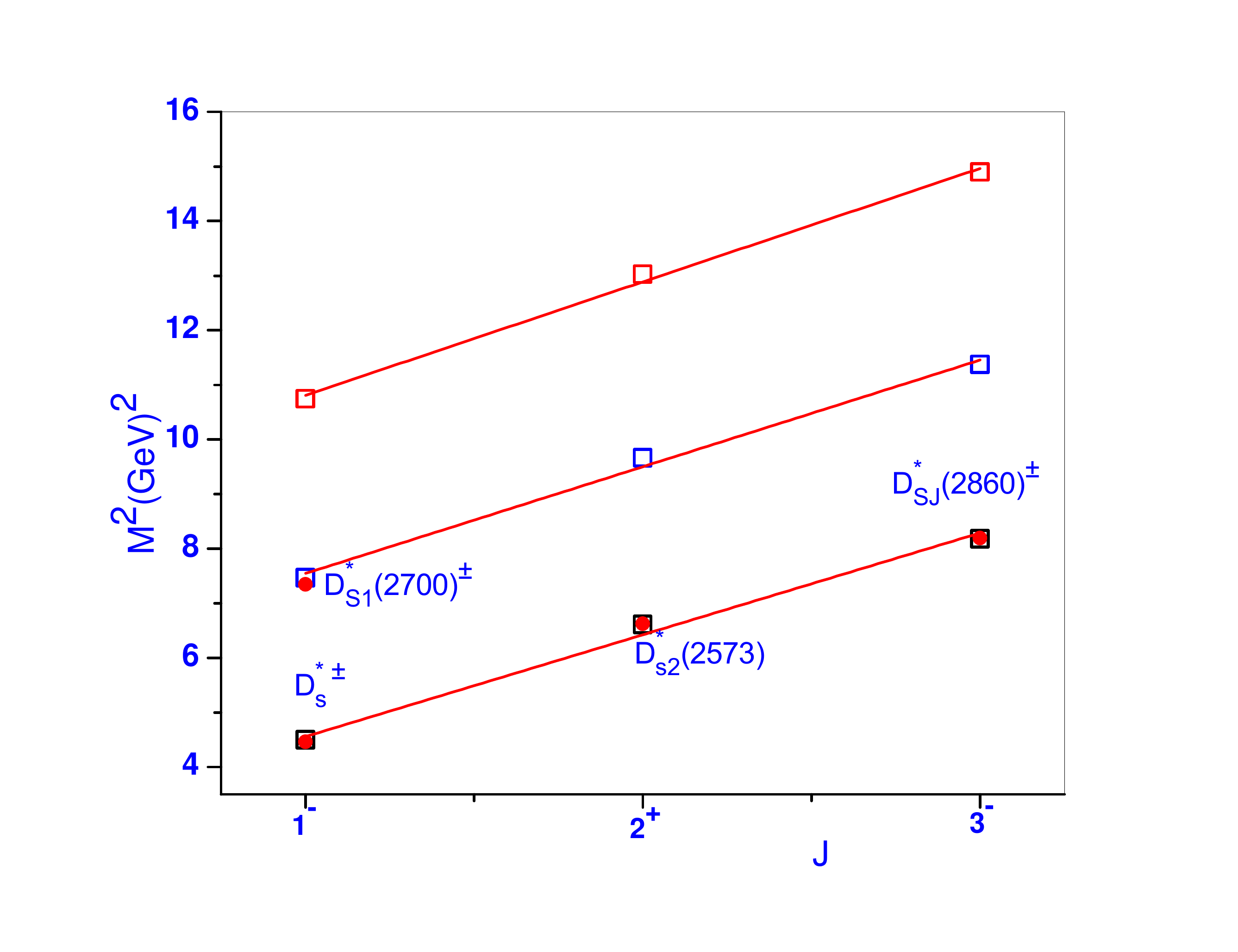}
    \figcaption{\label{fig:NPmesonDs}  Parent and daughter $(J,\: M^{2})$ Regge trajectories for charmed-strange mesons with natural parity. Hollow squares are predicted masses. Available experimental data are given by dots with particle names.}
      \end{center}
     
    \begin{center}
    \includegraphics[bb=30bp 40bp 750bp 575bp,clip,width=0.72
    \textwidth]{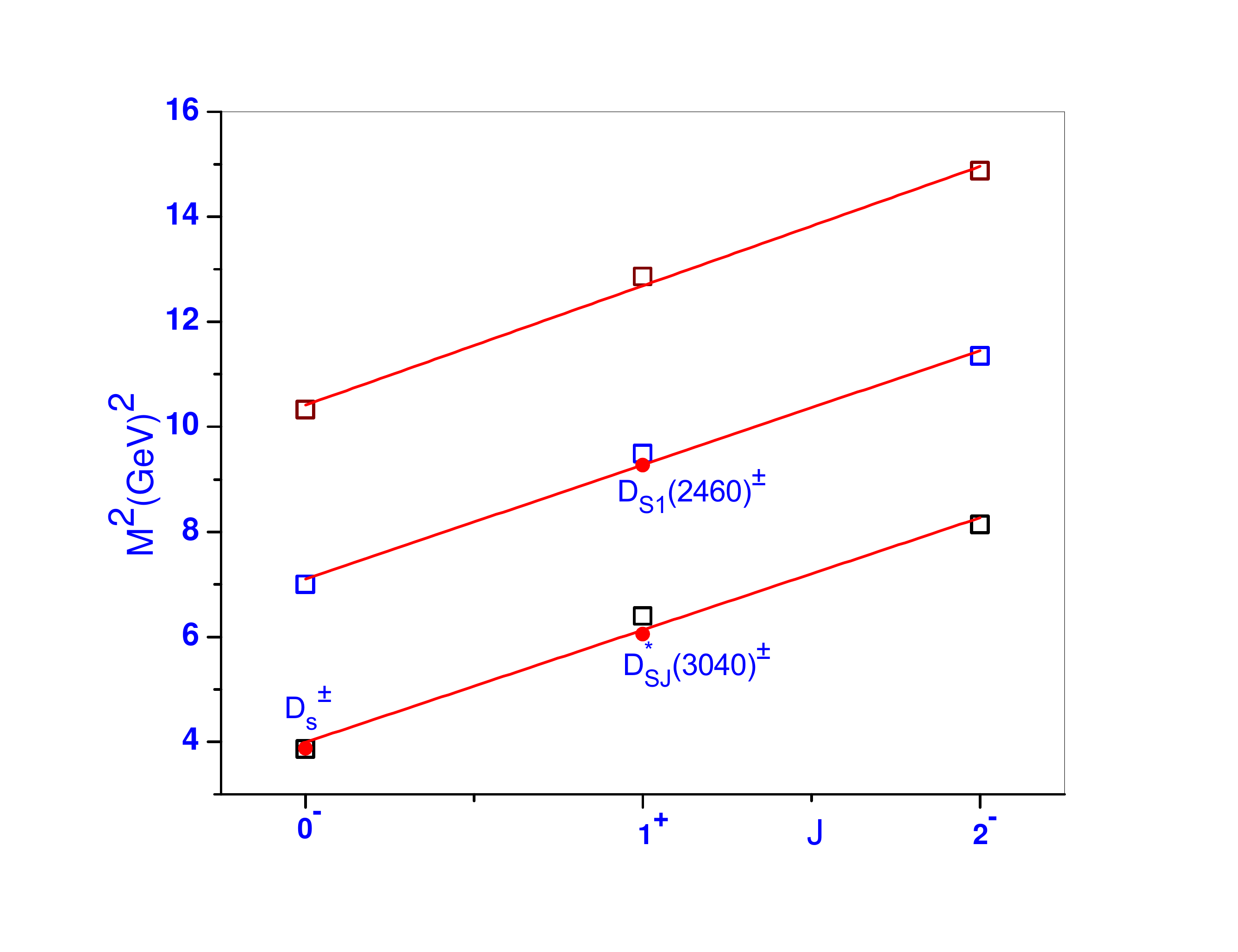}
    \figcaption{\label{fig:UNPmesonDs}  Parent and daughter $(J,\: M^{2})$ Regge trajectories for charmed-strange mesons with unnatural parity. Hollow squares are predicted masses. Available experimental data are given by dots with particle names.}
    \end{center}
     
   \begin{center}
   \includegraphics[bb=30bp 40bp 750bp 575bp,clip,width=0.72
    \textwidth]{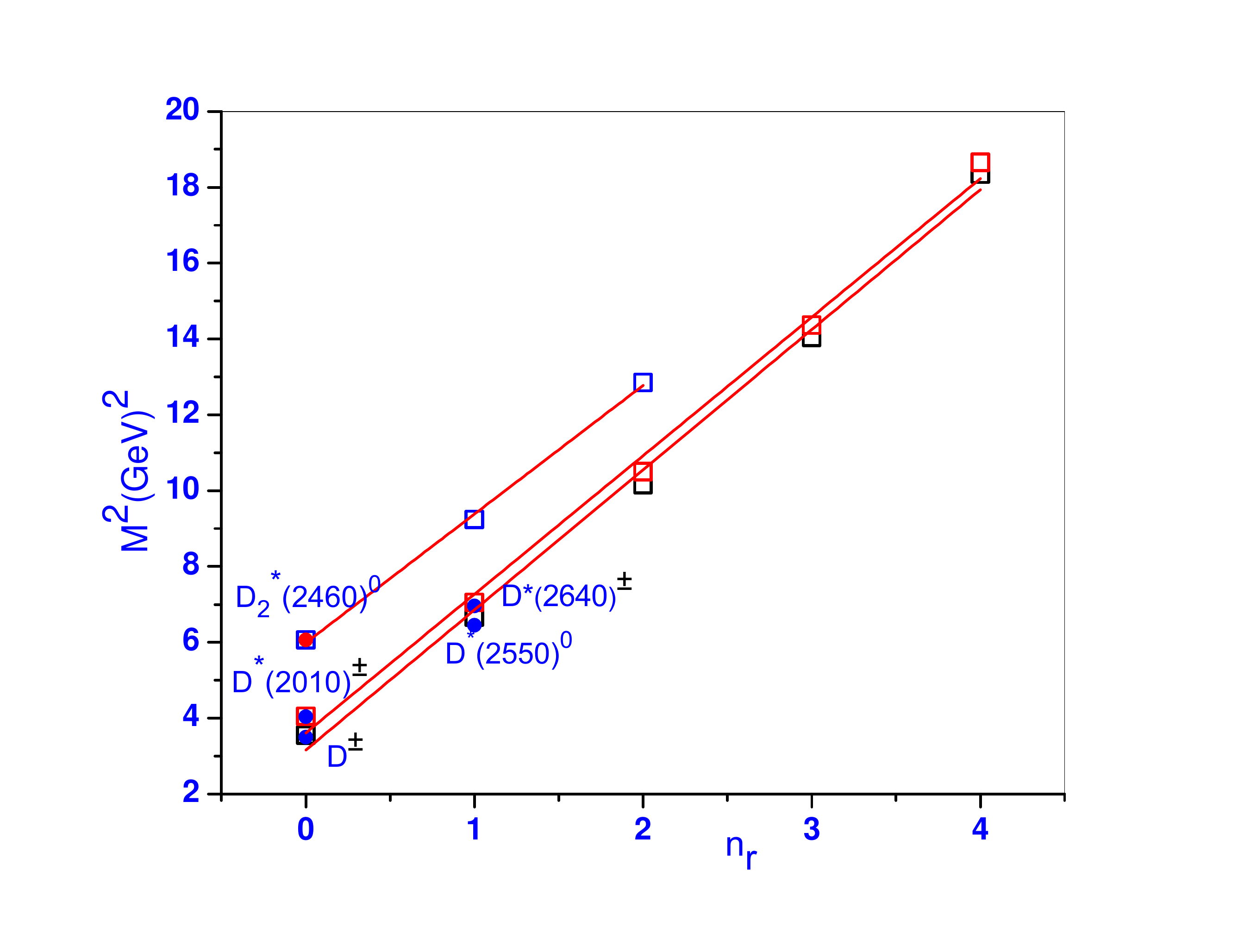}
    \figcaption{\label{fig:PsVmesonD} The $(n_{r},\: M^{2})$ Regge trajectories for pseudoscalar, vector and tensor charmed mesons (from bottom to top). Hollow squares are predicted masses. Available experimental data are given by dots with particle names.}
    \end{center}

   \begin{center}
   \includegraphics[bb=30bp 40bp 750bp 575bp,clip,width=0.72
    \textwidth]{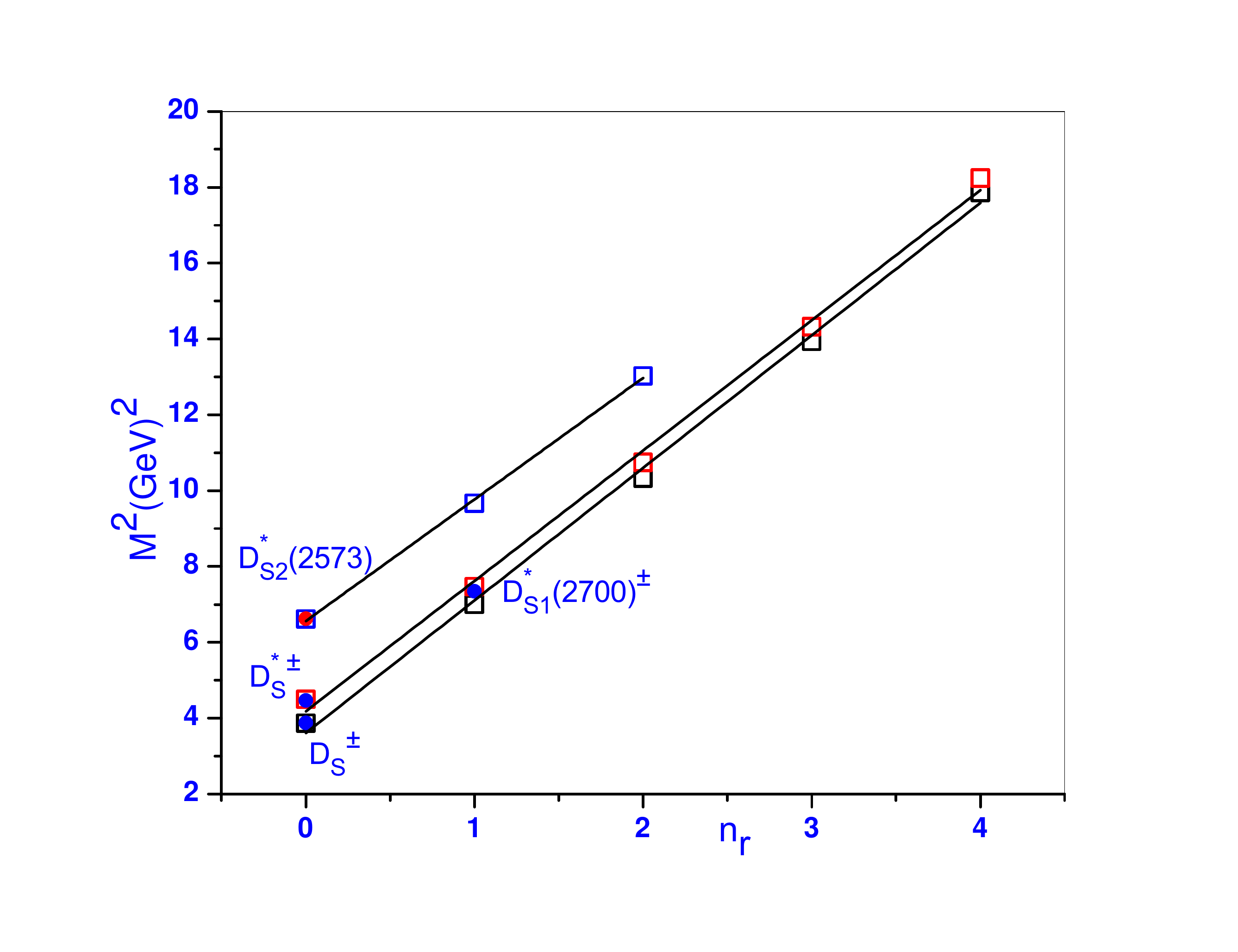}
  \figcaption{\label{fig:PsVmesonDs} The $(n_{r},\: M^{2})$ Regge trajectories for pseudoscalar, vector and tensor charmed-strange mesons (from bottom to top). Hollow squares are predicted masses. Available experimental data are given by dots with particle names. }
   \end{center}

  \begin{center}
  \includegraphics[bb=30bp 40bp 750bp 575bp,clip,width=0.72
  \textwidth]{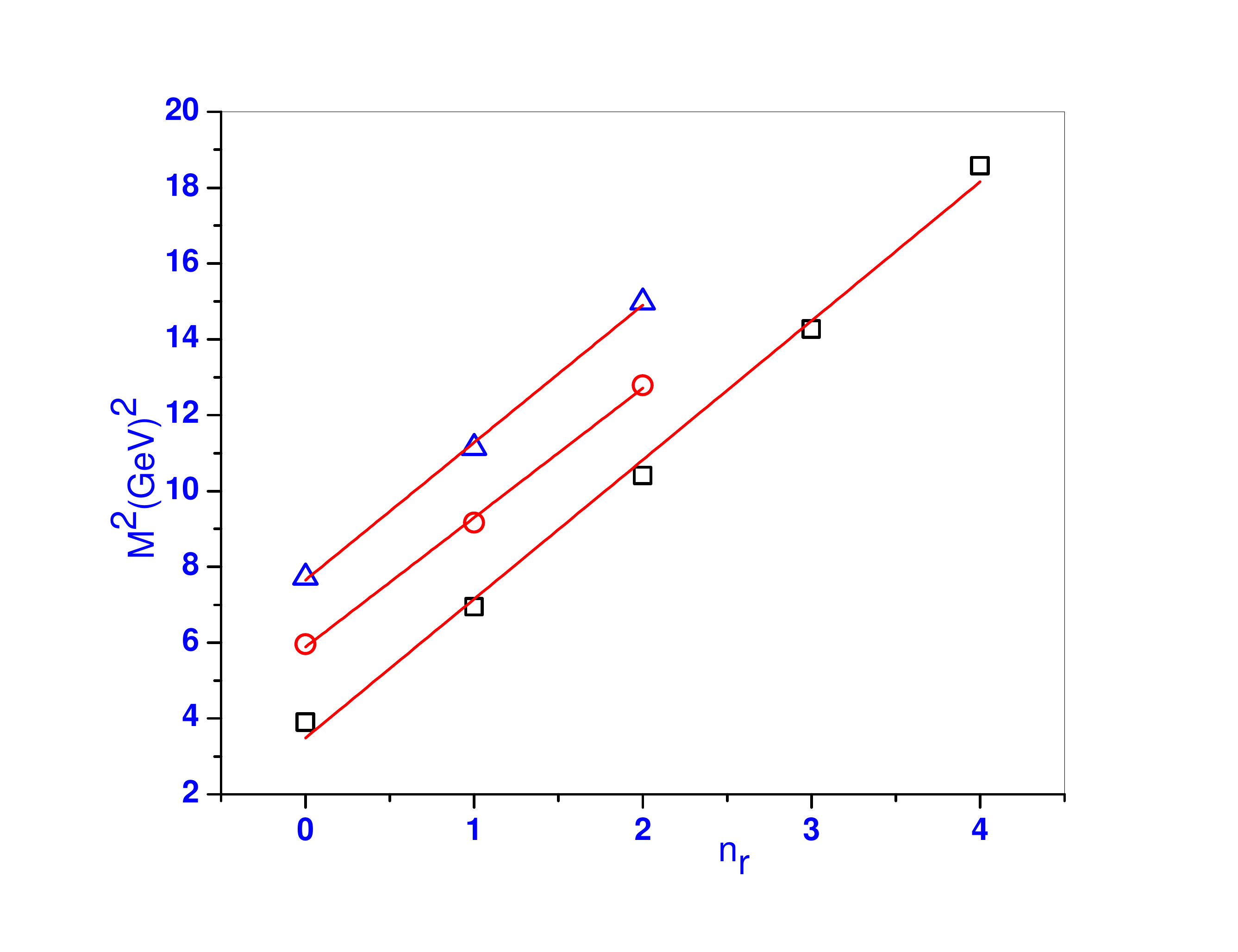}
  \figcaption{\label{fig:SavmesonD} The $(n_{r},M^{2})$ spin-average mass Regge trajectories of the S-P-D states (from bottom to top) for charmed mesons. }
   \end{center}
     
  \begin{center}
  \includegraphics[bb=30bp 40bp 750bp 575bp,clip,width=0.72
  \textwidth]{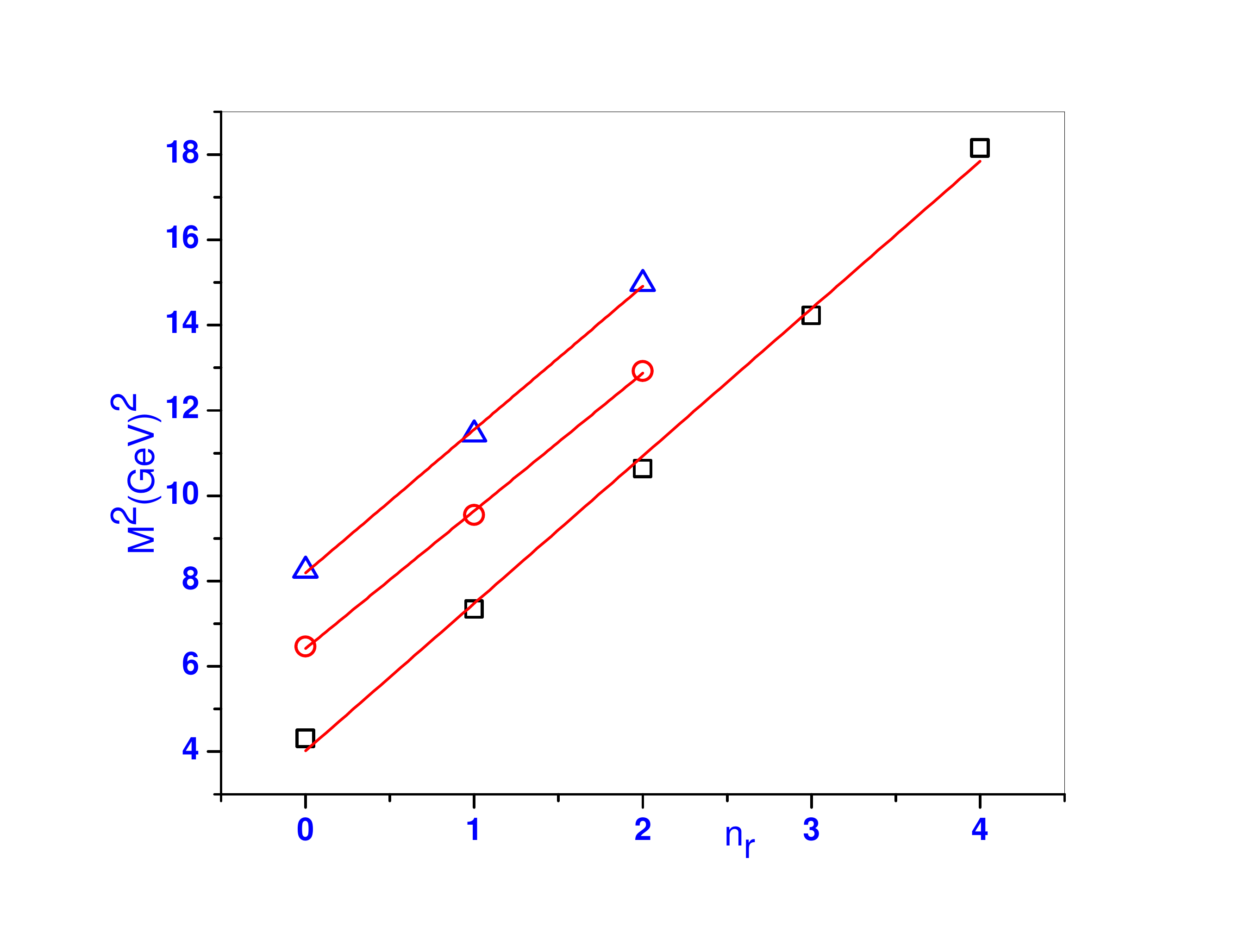}
  \figcaption{\label{fig:SavmesonDs} The $(n_{r},M^{2})$  spin-average mass Regge trajectories of the S-P-D states (from bottom to top) for charmed-strange mesons. }
  \end{center}

   \begin{table*}
   \caption{Fitted parameters of the $(J,\: M^{2})$ parent and daughter Regge trajectories for heavy-light mesons with unnatural and natural parity $(q=u/d)$.\label{tab:alfa}}
   \noindent \centering{}%
   \begin{tabular}{ccccc}
   \hline 
   \noalign{\smallskip}
   \multirow{2}{*}{Parity} & \multirow{2}{*}{Meson} & \multirow{2}{*}{Trajectory} & \multicolumn{1}{c}{\multirow{2}{*}{$\alpha(GeV^{-2})$}} & \multirow{2}{*}{$\alpha_{0}$}\tabularnewline\noalign{\smallskip}
   
   \noalign{\smallskip}\hline\noalign{\smallskip}
   \multirow{6}{*}{Unnatural} & \multirow{3}{*}{$D$$(c\bar{q})$} & Parent & $0.491\pm0.044$ & $-1.787\pm0.262$\tabularnewline
    &  & First daughter & $0.458\pm0.032$ & $-3.087\pm0.288$\tabularnewline
    &  & Second daughter & $0.424\pm0.023$ & $-4.333\pm0.293$\tabularnewline
   \noalign{\smallskip} 
   \noalign{\smallskip}
    & \multirow{3}{*}{$D_{s}$$(c\bar{s})$} & Parent & $0.462\pm0.049$ & $-1.834\pm0.315$\tabularnewline
    &  & First daughter & $0.456\pm0.039$ & $-3.233\pm0.366$\tabularnewline
    &  & Second daughter & $0.432\pm0.025$ & $-4.490\pm0.317$\tabularnewline
   \hline 
   \noalign{\smallskip}
   \multirow{6}{*}{Natural } & \multirow{3}{*}{$D^{*}$$(c\bar{q})$} & Parent & $0.527\pm0.019$ & $-1.151\pm0.112$\tabularnewline
    &  & First daughter & $0.475\pm0.011$ & $-2.363\pm0.103$\tabularnewline
    &  & Second daughter & $0.435\pm0.006$ & $-3.566\pm0.075$\tabularnewline
   \noalign{\smallskip}
   \noalign{\smallskip}
    & \multirow{3}{*}{ $D_{s}^{*}$$(c\bar{s})$} & Parent & $0.538\pm0.046$ & $-1.462\pm0.305$\tabularnewline
    &  & First daughter & $0.509\pm0.037$ & $-2.835\pm0.355$\tabularnewline
    &  & Second daughter & $0.480\pm0.027$ & $-4.182\pm0.357$\tabularnewline
   \hline 
   \end{tabular}
   \end{table*}

 \begin{table*}
 \caption{Fitted parameters of the $(n_{r},\: M^{2})$ Regge trajectories for heavy-light charmed mesons $(q=u/d)$ \label{tab:bita}}
 
 \noindent \begin{centering}
 \begin{tabular}{ccc}
 \hline \noalign{\smallskip}
 Mesons & $\beta(GeV^{-2})$ & $\beta_{0}$\tabularnewline
 \noalign{\smallskip} 
 \hline 
 \noalign{\smallskip}
 $c\bar{q}$ &  & \tabularnewline
 $D$ & $0.270\pm0.010$ & $-0.842\pm0.116$\tabularnewline
 $D^{*}$ & $0.272\pm0.011$ & $-0.974\pm0.131$\tabularnewline
 $D_{0}^{*}$ & $0.287\pm0.008$ & $-1.580\pm0.079$\tabularnewline
 $D_{1}$ & $0.292\pm0.010$ & $-1.690\pm0.096$\tabularnewline
 $D_{1}$ & $0.292\pm0.010$ & $-1.727\pm0.096$\tabularnewline
 $D_{2}$ & $0.294\pm0.011$ & $-1.750\pm0.105$\tabularnewline
 \hline 
 \noalign{\smallskip} 
 $c\bar{s}$ &  & \tabularnewline
 $D_{s}$ & $0.286\pm0.010$ & $-1.025\pm0.087$\tabularnewline
 $D_{s}^{*}$ & $0.290\pm0.009$ & $-1.207\pm0.110$\tabularnewline
 $D_{s0}^{*}$ & $0.303\pm0.006$ & $-1.790\pm0.054$\tabularnewline
 $D_{s1}$ & $0.309\pm0.008$ & $-1.965\pm0.077$\tabularnewline
 $D_{s1}$ & $0.308\pm0.007$ & $-1.978\pm0.074$\tabularnewline
 $D_{s2}$ & $0.312\pm0.009$ & $-2.043\pm0.087$\tabularnewline
 \hline 
 \end{tabular}
 \par\end{centering}
 \end{table*}

 \begin{table*}
 \caption{Fitted parameters of the $(n_{r},\: M^{2})$ S, P and D state spin-average mass Regge trajectories for heavy-light mesons $(q=u/d)$.\label{tab:Spinave} }
 
 \noindent \centering{}%
 \begin{tabular}{cccc}
 \hline\noalign{\smallskip}
 
 Meson & Trajectory & $\beta(GeV^{-2})$ & $\beta_{0}$\tabularnewline
 
 \noalign{\smallskip}\hline\noalign{\smallskip}
 
 \multirow{3}{*}{$D$$(c\bar{q})$} & S Satate & $0.271\pm0.010$ & $-0.937\pm0.126$\tabularnewline
  & P State & $0.292\pm0.010$ & $-1.720\pm0.098$ \tabularnewline
  & D State & $0.275\pm0.009$ & $-2.106\pm0.106$\tabularnewline
  
 \noalign{\smallskip}
 \noalign{\smallskip}
 \multirow{3}{*}{$D_{s}$$(c\bar{s})$} & S Satate & $0.298\pm0.005$ & $-1.225\pm0.062$\tabularnewline
  & P State & $0.305\pm0.005$ & $-1.930\pm0.049$\tabularnewline
  & D State & $0.297\pm0.008$ & $-2.432\pm0.098$\tabularnewline
 \hline 
 \medskip
 \end{tabular}
 \end{table*}

 \begin{multicols}{2}

\section{Conclusion\label{sec:conclusion}}
The mass spectra of $D$ and $D_S$ mesons have been calculated in a semi-relativistic approach by using a phenomenological quark-antiquark color-Coulomb plus linear potential. The kinetic energy term includes relativistic corrections,  and the potential energy term incorporates  first-order relativistic corrections. The spin-averaged masses of both the mesons have been calculated (see Table (\ref{tab:swavespin})).  The mass spectra given in Tables (\ref{tab:massesD}) and (\ref{tab:massesDs}) are close to the experimental values and other theoretical estimates. 

The ${1}^3P_{0}$ states of the $D$ and $D_S$ mesons are overestimated by 46 MeV and 120 MeV respectively. We found a similar trend for other theoretical estimates. Lattice QCD estimates for the $D$ meson $1^{3}P_{0}$ state are in excellent agreement with a difference of 7 MeV, but in the case of the $D_S$ meson they are overestimated by 73 MeV \cite{Cichy2016}. In Ref.~\cite{Godfrey2015}, the ${1}^3P_{0}$ states are  overestimated by 81 MeV and 166 MeV, and in Refs.~\cite{Shah2014,Shah2016} by 93 MeV and 31 MeV respectively for the $D$ and $D_S$ mesons. Masses predicted by all the models are overestimated, which suggests that these two experimental states may be good candidates for exotic states. We have estimated the effects of relativistic corrections. In the case of masses of S states, the maximum relativistic correction is found to be about 9\%, in the case of the P states it is about 15\%, and in the case of the D states it is about 13\%. Overall, a slight improvement is observed on our previous calculation in Ref.~\cite{Devlani2011,Devlani2013}, which shows the importance of the relativistic corrections in kinetic as well as in potential energy.

Leptonic branching fractions have been evaluated using spectroscopic parameters of these mesons and are tabulated in table (\ref{tab:leptobranch}). The obtained results are compared with PDG \cite{PDGlatest} values. The leptonic branching fractions are slightly underestimated for the $D$ and $D_S$ mesons.  The radiative leptonic decay widths and branching ratios are tabulated in Table ~(\ref{tab:radiativelepto}). In the absence of experimental  measurements, we have compared our results with those of Ref.~\cite{Cai-Dian2003}, but the results are not in mutual agreement. Using the predicted results we have estimated the mixing parameters of the $D$ meson. The values of $\Delta m_{q}$, $x$, $y$, $\chi$ and $R_M$ are tabulated in Table (\ref{tab:mixing}), which shows reasonably close agreement with experimental values. 
 
 Finally, we plotted the Regge trajectories in the $(J,M^{2})$ plane for these charmed and charmed strange mesons with natural $(P=(-1)^{J})$ and unnatural $(P=(-1)^{J-1})$ parity. We also plotted the Regge trajectories in the $(n_r, \ M^{2})$ plane for spin-averaged masses as well as for pseudoscalar, vector and tensor charmed and charmed strange mesons.  $\chi^{2}$ fitting was done for the slope and intercepts of these Regge trajectories. We see that the calculated heavy-light meson masses fit nicely to the linear trajectories in $(J,M^{2})$ and  $(n_r, \ M^{2})$ planes with almost parallel and equidistant lines. These Regge trajectories could play a very important role in identifying any new excited state (experimentally), as they provide information about the quantum numbers of the particular state.   

This study has shown the importance of relativistic correction within the potential model for successful prediction of the spectroscopic parameters for the $D$ and $D_S$ mesons.\\

{\bf Acknowledgements}

A. K. Rai acknowledge the financial support extended by Department of Science of Technology, India  under SERB fast track scheme SR/FTP /PS-152/2012 and also to SVNIT (Institute Research Grant (Dean ($R\& C$)/1488/2013-14)).\\

\bibliographystyle{epj}
\bibliography{myref}

\end{multicols}

\vspace{-1mm}
\centerline{\rule{80mm}{0.1pt}}
\vspace{2mm}

\begin{multicols}{2}

\end{multicols}

\clearpage

\end{document}